\documentclass[11pt,draft]{IEEEtran}

\usepackage{color}     
\usepackage{epsf,psfrag,amssymb,amsfonts,latexsym,cite,verbatim,enumerate}
\usepackage{srcltx,amsmath,cases, graphicx}
\usepackage{times, multirow, array}
\usepackage[mathscr]{eucal}

\def\psfancypar#1#2{\begingroup\def\par{\endgraf\endgroup\lineskiplimit=0pt}
               \setbox2=\hbox{\large\sc #2}
               \newdimen\tmpht \tmpht \ht2 \advance\tmpht by \baselineskip
               \font\hhuge=Times-Bold at \tmpht
               \setbox1=\hbox{{\hhuge #1}}
               \count7=\tmpht \count8=\ht1
               \divide\count8 by 1000 \divide\count7 by \count8 
               \tmpht=.001\tmpht\multiply\tmpht by \count7 
               \font\hhuge=Times-Bold at \tmpht
               \setbox1=\hbox{{\hhuge #1}}
               \noindent
                \hangindent1.05\wd1
               \hangafter=-2 {\hskip-\hangindent
               \lower1\ht1\hbox{\raise1.0\ht2\copy1}%
                \kern-0\wd1}\copy2\lineskiplimit=-1000pt}

\def\thetabf{{\mbox{\boldmath$\theta$\unboldmath}}}

\newcommand{\Phibf}{\mbox{${\bf \Phi}$}}
\newcommand{\Gammabf}{\mbox{${\bf \Gamma}$}}

\newcommand{\E}{\mbox{{\rm E}}}

\newcommand{\abf}{\mbox{${\bf a}$}}

\def\boxit#1{\vbox{\hrule\hbox{\vrule\kern3pt
        \vbox{\kern3pt#1\kern3pt}\kern3pt\vrule}\hrule}}

\def\reals{ { {\rm  I \kern-0.15em R }  } }
\def\complex{ {\,{{\rm C} \kern-0.50em \raise0.20ex {  |}}\, }}

\def\mubf{\hbox{\boldmath$\mu$\unboldmath}}

\def\phibf{\hbox{\boldmath$\phi$\unboldmath}}

\def\Sigmabf{\hbox{$\bf \Sigma$}}

\def\Gammabf{\hbox{$\bf \Gamma$}}
\def\Thetabf{\hbox{$\bf \Theta$}}
\def\Lambdabf{\mbox{$ \bf \Lambda $}}

\def\Pibf{{\bf \Pi}}
\def\abf{{\bf a}}

\def\cbf{{\bf c}}

\def\nbf{{\bf n}}

\def\qbf{{\bf q}}
\def\rbf{{\bf r}}
\def\sbf{{\bf s}}

\def\ubf{{\bf u}}
\def\vbf{{\bf v}}
\def\wbf{{\bf w}}
\def\xbf{{\bf x}}
\def\ybf{{\bf y}}
\def\zbf{{\bf z}}
\def\rbf{{\bf r}}
\def\xbf{{\bf x}}
\def\ybf{{\bf y}}
\def\Abf{{\bf A}}
\def\Bbf{{\bf B}}
\def\Cbf{{\bf C}}
\def\Dbf{{\bf D}}

\def\Hbf{{\bf H}}
\def\Ibf{{\bf I}}

\def\Mbf{{\bf M}}

\def\Pbf{{\bf P}}
\def\Qbf{{\bf Q}}
\def\Rbf{{\bf R}}

\def\Ubf{{\bf U}}
\def\Vbf{{\bf V}}

\def\Xbf{{\bf X}}

\def\Zbf{{\bf Z}}

\def\Cc{{\cal C}}

\def\Nc{{\cal N}}

\def\Sc{{\cal S}}

\def\be{\vskip .3cm \begin{equation}}
\def\ee{\end{equation} \vskip .4cm \noindent}
\def\defeq{{\stackrel{\Delta}{=}}}
\newcommand{\R}{\mbox{$\hat {\bf R}_{N}$}}

\def\Rxx{\Rbf_{\ssstyle X\kern-.1em X}}

\let\ssstyle=\scriptscriptstyle

\def\Kout{\setbox1=\hbox{\Huge\bf K}\hbox to
1.05\wd1{\hspace{.05\wd1}
}}
\def\Sout{\setbox1=\hbox{\Huge\bf S}\hbox to 1.05\wd1{\hspace{.05\wd1}
}}

  \ifx\LabelFigloaded\MYundefined\relax
  \else
   \fi

  \def\LabelFigloaded{\relax}

  \chardef\LabelFigCatAt\the\catcode`\@
  \catcode`\@=11

 \let\LabelFigwlog@ld\wlog
 \def\wlog#1{\relax}

 \ifx\\\MYundefined@
    \let\\\relax
 \fi

  \def\ms@g{\immediate\write16}

 \def\N@wif{\csname newif\endcsname }
 \def\Temp@ {\N@wif\ifIN@}
 \ifx\INN@\MYundefined@
    \else \let\Temp@\relax
 \fi
 \Temp@

  \def\IN@{\expandafter\INN@\expandafter}
  \long\def\INN@0#1@#2@{\long\def\NI@##1#1##2##3\ENDNI@
    {\ifx\m@rker##2\IN@false\else\IN@true\fi}%
     \expandafter\NI@#2@@#1\m@rker\ENDNI@}
  \def\m@rker{\m@@rker}
 
  \newtoks\Initialtoks@  \newtoks\Terminaltoks@
  \def\SPLIT@{\expandafter\SPLITT@\expandafter}
  \def\SPLITT@0#1@#2@{\def\TTILPS@##1#1##2@{%
     \Initialtoks@{##1}\Terminaltoks@{##2}}\expandafter\TTILPS@#2@}

 \def\Shifted@@#1#2#3{\setbox0=\hbox{#3}%
   \raise -\dp0\vbox {\kern-#2%
       \hbox {\kern#1\unhbox0\kern-#1}%
           \kern#2}}

 \newcount\gridcount
 \newbox\auxGridbox@ \newbox\hGridbox@ \newbox\vGridbox@
 \newbox\Labelbox@ \newbox\auxLabelbox@
 \newbox\Coordinatebox@
 \newtoks\Labeltoks@
 \newdimen\Wdd@ \newdimen\Htt@
 \newdimen\Wddd@ \newdimen\Httt@
 
 \def\Wr@{\immediate\write16}

 \newdimen\GL@wd
 \def\GridLineWidth#1{\GL@wd=#1}

 \def\gobble#1{}
 \def\EdgeErr@{\Wr@{}%
      \Wr@{\string\Edges\space argument
      1, 10, 100 or 1000 please\string!}%
      }

 \newcount\Edgect@

 \def\Sweepup#1\endSweepup{}

 \def\SetEdges@{%
    \edef\Zr@@s{\expandafter\gobble\number\Edgect@\empty}%
        \count255=0\Zr@@s\relax
        \ifnum\count255=\z@\else\EdgeErr@\show\tailtest\fi
        \count255=1\Zr@@s\relax
        \ifnum\count255=\Edgect@\relax\else\EdgeErr@\show\leadtest\fi
    \EdgGl@b\edef\Zr@s{\expandafter\gobble\Zr@@s\empty}
    \ifnum\Edgect@>\@ne\relax\EdgGl@b\let\L@Dc\empty
        \else\EdgGl@b\edef\L@Dc{\string.}\fi
    \ifnum\Edgect@>\@ne\relax
        \EdgGl@b\edef\Edgescale@##1{\divide##1 by \Edgect@}%
        \else\EdgGl@b\edef\Edgescale@##1{}\fi
    }

 \def\Edges#1{\Edgect@=#1\relax
     \let\EdgGl@b\global \SetEdges@}

 \Edges{1}

 \def\hhrule{\hrule height \GL@wd\vskip-.\GL@wd}

 \def\hRule@{%
   \advance\gridcount -2%
   \vfil\hhrule\vfil
   \llap{\smash{\raise -2.5pt
     \hbox{\L@Dc\number\gridcount\Zr@s\kern2pt}}}%
   \hhrule
   }

\def\vvrule{\vrule width \GL@wd \kern-\GL@wd}

 \def\vRule@{\advance\gridcount 2%
   \hfil\vvrule\hfil
   \setbox\auxGridbox@=\vbox to 0pt
      {\vskip \Htt@\vskip 2pt
        \hbox to 0pt{\hss\L@Dc\number\gridcount\Zr@s\hss}\vss}%
      \wd\auxGridbox@=0pt \box\auxGridbox@
   \vvrule
   }

 \def\PlaceGrid@@{\gridcount=10 
  \setbox\hGridbox@=\hbox{%
        \hbox{%
             \hskip-.4pt\vrule
             \vbox to \Htt@{%
               \offinterlineskip\parindent=\z@\relax
               \hbox to \Wdd@{\hfil}
               \hRule@\hRule@\hRule@\hRule@
               \vfil\hhrule\vfil}%
             \vrule\hskip-.4pt}
    }%
  \gridcount=0%
  \setbox\vGridbox@=\hbox{%
      \vbox{\offinterlineskip\parindent=0pt\hsize=0pt
         \vskip-.4pt\hrule%
         \hbox to \Wdd@{%
                 \vtop to \Htt@{\vfil}%
                 \vRule@\vRule@\vRule@\vRule@
                 \hfil\vvrule\hfil}%
         \hrule\vskip-.4pt}}%
  \wd\hGridbox@=0pt\ht\hGridbox@=0pt
  \wd\vGridbox@=0pt\ht\vGridbox@=0pt
  \hbox{\box\hGridbox@\box\vGridbox@}%
  }

 \def\LabelsGlobal{\def\LabGl@b{\global}}
 \def\LabelsLocal{\def\LabGl@b{}}
 \LabelsGlobal 

 \def\SetLabels#1\endSetLabels{%
   \LabGl@b\Labeltoks@={#1()\\}%
   }

 \LabGl@b\Labeltoks@={()\\}

 \def\ShowGrid{\LabGl@b\let\PlaceGrid@\PlaceGrid@@}
 \def\HideGrid{\LabGl@b\let\PlaceGrid@\relax}
 \def\Grids{\ShowGrid\LabGl@b\let\GridSwitch@\ShowGrid}
 \def\noGrids{\HideGrid\LabGl@b\let\GridSwitch@\HideGrid}

 \noGrids

 \def\bAdjust@@{%
     \setbox\auxLabelbox@=\hbox{\raise \dp\auxLabelbox@
            \box\auxLabelbox@}}
 \def\bAdjust@{\let\vAdjust@\bAdjust@@}

 \def\eAdjust@@{\dimen0=-.5\ht\auxLabelbox@
     \advance\dimen0 by .5\dp\auxLabelbox@
     \setbox\auxLabelbox@=
            \hbox{\raise\dimen0\box\auxLabelbox@}}
 \def\eAdjust@{\let\vAdjust@\eAdjust@@}

 \def\tAdjust@@{%
     \setbox\auxLabelbox@=\hbox{\raise-\ht\auxLabelbox@
            \box\auxLabelbox@}}
 \def\tAdjust@{\let\vAdjust@\tAdjust@@}

 \let\vAdjust@\relax

 \def\lAdjust@{\let\hAdjust@\rlap}
 \def\rAdjust@{\let\hAdjust@\llap}

 \let\hAdjust@\relax\let\vAdjust@\relax

 \def\FetchLabel@#1(#2)#3\\{%
     \IN@0#2@@\ifIN@
        \setbox0=\hbox{\ignorespaces#1#3\unskip}%
        \ifdim\wd0>0pt
           \ms@g{}%
           \ms@g{ !!! Bad label(s)? !!!}%
           \message{ #1(#2)#3}%
        \fi
        \def\LabelMole@##1\endFetchLabel@{%
            \IN@0()\\@##1@%
            \ifIN@\def\Temp@{\FetchLabel@##1\endFetchLabel@}%
            \else\def\Temp@{}%
            \fi
            \Temp@
           }%
     \else
       \ignorespaces#1\unskip
       \setbox\auxLabelbox@=%
         \hbox to 0pt{\hss\ignorespaces\hAdjust@
          {\ignorespaces#3\unskip}\hss}%
       \vAdjust@
       \let\hAdjust@\relax\let\vAdjust@\relax
       \AugmentLabelBox@@{#2}%
       \ht\Labelbox@=0pt\dp\Labelbox@=0pt
       \let\LabelMole@\FetchLabel@%
     \fi\LabelMole@}

 \newtoks\XYSep@ 
 \def\SetXYSeparator#1{%
     \IN@0#1@@\ifIN@\XYSep@{*}%
     \else
     \XYSep@{#1}%
     \fi
     }

 \SetXYSeparator*

 \def\AugmentLabelBox@@#1{%
     \IN@0\the\XYSep@ @#1@\ifIN@
       \SPLIT@0\the\XYSep@ @#1@%
       \setbox\Labelbox@=\hbox to 0pt{%
         \unhbox\Labelbox@
         \Shifted@@{\the\Initialtoks@\Wddd@}%
         {\the\Terminaltoks@\Httt@}%
         {\box\auxLabelbox@}}%
     \else
         \ms@g{}%
         \ms@g{ !!! Bad insertion point. !!!}%
         \message{ (#1\ this point was rejected.)}%
     \fi
    }

 \def\FetchOption@#1[#2]#3\endFetchOption@{%
    \def\temp{#1}
    \ifx\temp\empty
       \Edgect@=#2\relax
       \let\EdgGl@b\relax
       \SetEdges@
       \Cleaner@#3%
    \fi}

 \def\Cleaner@#1[@]{\Labeltoks@{#1}}
     
 \def\PlaceLabels@@{\mathsurround=0pt
     \def\Cr@{\\}%
     \let\L\lAdjust@\let\R\rAdjust@
     \let\B\bAdjust@\let\E\eAdjust@\let\T\tAdjust@
     \expandafter\FetchOption@\the\Labeltoks@[@]\endFetchOption@
     \Wddd@=\Wdd@ \Edgescale@\Wddd@ 
     \Httt@=\Htt@ \Edgescale@\Httt@
     \expandafter\FetchLabel@\the\Labeltoks@\endFetchLabel@
     \box\Labelbox@
     }%

 \let \PlaceLabels@\PlaceLabels@@

 \def\AffixLabels#1{\setbox\Coordinatebox@=\hbox{#1}%
      \Wdd@=\wd\Coordinatebox@ \Htt@=\ht\Coordinatebox@
      \advance\Htt@ \dp\Coordinatebox@
      \hbox{\copy\Coordinatebox@\kern-\Wdd@ 
           \Shifted@@{0pt}{-\dp\Coordinatebox@}%
           {\PlaceLabels@\PlaceGrid@}%
           \kern\Wdd@}%
      \GridSwitch@ 
      \LabGl@b\Labeltoks@{()\\}%
      }
 
   \let\wlog\LabelFigwlog@ld   
   \catcode`\@=\LabelFigCatAt  

                                By

              Raymond S\'eroul <A18645@FRCCSC21.BITNET>
                                and 
              Laurent Siebenmann <lcs@topo.math.u-psud.fr>
    
              VERSIONS: July 1991, Oct 1991, Jan 1992, July 1992

INTRODUCTION

      This labelling package is intended for TeX users who
rely on non-TeX sources for for their graphics inserts.  It
provides means for adding TeX labels to such inserts with a
minimum of fuss. 

       For most labels, TeX users have in the past found it
reasonably convenient to rely on non-TeX sources. Typical
occasions when an inescapable need for TeX labels seemed to
arise are

 (a) when the graphics program lacks certain exotic or complex
mathematical symbols

 (b) when the very highest typographical quality is wanted for the
labels

 (c) when labels included with the graphics fail to print, 
 and you cannot figure out why (cf. boxedeps.doc).  The labels
 provided by labelfig.tex are 100

       Since this package first appeared, many users, who in the
past scarcely dreamed of using TeX labels, have come to use
nothing but.  So it is now appropriate to add

Intoxication Warning:  TeX labels may be addictive and expensive. 

     If you have a fast preview you may disagree, and even find
that this package provides an agreeable paste-up environment; see
extra applications at end.

     Note to publishers: It is possible and convenient to ultimately
export the TeX labels produced by labelfig.tex to become an integral
part of the EPS file. This is often desired by a publisher who typically
uses an "upmarket" graphics or page layout program, with which the
staff is skilled in perfecting figures.  See Appendix I for
a recipe.

     The authors are grateful to Patrick Ion of Math Reviews for
helpful comments and encouragement.

BASIC INSTRUCTIONS

    After reading in the macro file using

preview or proof your figure with a coordinate grid printed on
top, by typing the following:

    \ShowGrid  
    \AffixLabels{<the graphics insertion>}

Here <the graphics insertion> is what you would type to insert
the graphics object alone without the grid.  This must provide
for the space around it. For example <the graphics insertion>
might well be \BoxedEPSF{MyFigure scaled 700} using the
boxedeps.tex macro package (from same source); this provides a
TeX box containing the encapsulated PostScript insert specified by
the file MyFigure. \AffixLabels{...} provides the grid (supposing
\ShowGrid is present) and later, once you have specified labels
using the grid, it will "tack on" the labels.

     The grid is a sort of (usually elongated) checkerboard of
ten rows and ten columns and its (internal) partitions are by
default numbered  .1, ... ,.9  both horizontally (X-coordinate
running left to right) and vertically (Y-coordinate running bottom
to top).  Thus the points enclosed by the grid correspond to the
points of the unit square in the cartesian "X-Y" plane, the lower
left corner corresponding to the origin (0,0).  By extrapolation,
the full page corresponds to a larger rectangle in the plane.

     These coordinates serve to position labels as follows.
Before the \AffixLabels{...} command type label specifications:

  \SetLabels
   (<X-coordinate>*<Y-coordinate>) <first label> \\
   .
   .
   .
   (<X-coordinate>*<Y-coordinate>)  <last label> \\
  \endSetLabels

Each row specifies one label and is terminated by \\.  In each
row, the position indicator comes first; it is written as a
standard cartesian point except that the X- and Y- coordinates
are separated by * rather than a comma because TeX allows a
comma as decimal point. There are no dimension units to specify
as the unit is the grid itself.

     By default, this cartesian point specifies where the middle
of the baseline of the label will be located.  However if you precede
the point by \L [or \R] the left [or right] edge of the baseline will
be located there. Similarly you may also precede the point by \T, \E,
or \B to vertically align the top equator or bottom of the label box
at the specified point.  This gives nine standard positions of
the label with respect to the insertion point --- corresponding to
the eight principle points of the compas and the center

                     \L\T     \T      \R\T

                     \L\E     \E      \R\E

                     \L\B     \B      \R\B

But this neglects the default "baseline" level of TeX,
giving potentially three more positions

                     \L    <no tag>   \R

For text, the baseline level is often the preferred. Its relation to
the others is variable. It will often coincide with the bottom level,
as happens for "X".  But it is often distinct, as for "g", in which
case you have in all 12 distinct positions rather than 9.

     It is convenient to think of this specification of label
position as attaching the label by a thumb-tack to the coordinate
grid. There are up to twelve positions of the thumb-tack on the
label, while the position of the thumb-tack on the coordinate grid is
arbitrary.  Normally, one choses the position of the thumb-tack on
the label to be the one that is the closest to the item being
labeled.  There are good reasons for this "rule of thumb":

   (a)  It facilitates correct positioning at first try.

   (b)  If the scale of the figure must be altered after labels
have been affixed, the labels have a good chance of remaining well
positioned.

   (c)  The visible grid need not extend beyond the "bounding box"
for the figure, because the best preferred position is always
(at least almost) within the bounding box .

The second reason is particularly important. Indeed it often
happens that scale has to be altered after labelling begins, in
order to either provide space for the labels, or to adjust
proportions between the labels and the figure.  (The size of labels
is unaffected by scaling.)

     Here is an artificial but self-contained test which uses
TeX rules to make a graphics object.

TEST

    Do not skip this!

 \def\FrameIt#1{\hbox{\vrule$\vcenter {\hrule\kern3pt%
             \hbox {\kern3pt #1\kern3pt}%
               \kern3pt\hrule}$\relax\vrule}}

 \def\Caption#1#2{\FrameIt{%
       \vtop {\hsize=#1\relax \parindent=0pt
         \leftskip=0pt \rightskip=0pt plus15pt
         \parfillskip=0pt
         \lineskip=1pt\baselineskip=0pt
         #2}}}

 \def\FirstQuadrant{\hbox to 100pt{\vrule\vbox to 100pt{%
        \hbox to 100pt{\hfil}\vfil\hrule}\hss}}

  \SetLabels
    \R(.5*.2) $\zeta\,\cdot$\\
    (.9*-.10) $\xi$\\
    \R(-.03*.9) $\eta$\\
    \T(.5*.9) \Caption{70pt}{%
          \it The norm of
          $g(\xi+i\eta)$ is indicated on
          contours of this invisible surface.}\\
  \endSetLabels

  \AffixLabels{\FirstQuadrant}

  \end

  Note that the coordinates to use for labels are indicated on the
edges of the grid (when visible) corresponding to the conventional
x- and y- axes of the Cartesian plane. By default the grid is
1-by-1. However, by the command \Edges{100}, you can change this
to 100-by-100 and many users find this alternative most
convenient. Place the command \Edges{...} in your style file (or
header) since its effect is is global. Other possible edge values
are 10 and 1000.

  If you use the command \Edges{...} at all, do so with care.  For
if you accidentally delete an \Edges{...} command your labels will
abruptly be badly misplaced and may logically but mysteriously
generate "dimension too big" errors under TeX and "off page" errors
under your driver.  

  You can dictate the edgescale for an individual figure by giving
the scale in brackets immediately after \SetLabels.  Thus, to
import into an article using say \Edge{100} a figure labelled using
another edgescale, say the original 1-by-1 default, you can use
\SetLabels[1]...\endSetLabels.

GETTING IT DOWN PAT

     Complicated labeling deserves the same respect as
complicated mathematics.  Do not expect it to come out perfect the
first time!  What is needed in either case is a mechanism to
repeatedly typeset troublesome pieces.

     One mechanism is always available.  One does complicated
labelling in a separate "test" file involving just the figure being
labelled;  a texpert will know how to \dump TeX's current state as
a temporary format that restarts rapidly at each retry.  Usually,
one then pastes the completed labelled figure back into the main
TeX file, but, of course, one can also \input it as an auxiliary
file.

     If you do not have a TeXpert at handy, here is a first
approximation to an efficient setup. By deletions reduce a copy
of your article to just a few lines before and after the figure.
Now label the figure, and finally, copy and paste the labelled
figure to the original article. Then copy the next figure to label
into this testbed and repeat. The TeXpert can improve the  speed
at which TeX starts up, by compiling a format specifically for
your article; just one caution: best NOT include in the format
ephemeral details of setup like \Set<mydriver>ArtSpecials (from
boxedeps.tex because this reads  figure dimensions which you may
change during your work session.

     An improved mechanism to repeatedly typeset troublesome
pieces is now available on the Macintosh; it is called LinoTeX;
see the same ftp sources.  It could be set up on many types
of computer.

     Before using labelfig.tex to attach labels to a graphics
object inserted using boxedeps.tex or BoxedArt.tex, make it a
firm rule to carefully adjust the bounding box using the trimming
commands of these packages, and also at least tentatively scale
and position the object. Beware of changing the grid inadvertently
after the labels have been positioned.  For example, correcting
the bounding box of a PostScript graphics object can foul up the
labels by changing the coordinate grid to which the labels are
attached. This is particularly true for the trimming  commands of
boxedeps.tex and BoxedArt.tex. However, as noted already, change
of scale is much less disruptive, and modest adjustments should be
well tolerated.

     Sometimes the labels protrude so far from the bounding box
of a figure that the figure has to be repositioned.  Best do this
by ad hoc spacing, say using \hglue and \vglue; altering the
bounding box would create a vicious circle.

     Remember that you are responsible for preventing labels
from overlapping. You are responsible for all label typography
including size and style. A label is really just about anything
that can be put in a TeX box. Note that spaces at the beginning
and end of labels will normally be suppressed; if you really want
them you must protect them with TeX braces.

     This package temporarily sets the \mathsurround parameter
of TeX to zero  while the labels are being affixed. This is done
because nonzero \mathsurround space would influence the position
of left and right aligned labels; then, when a texpert or printer
modifies mathsurround, diagram labeling might be disastrously
altered. There is a small price to pay involving labels that are
formatted as caption boxes including mathematics: you  may want or
need to specify an explicit mathsurround space within the caption
box; it will not influence anything outside.

     Those hostile to the use of * as separator between
the X and Y coordinates of label insertion points, are free to
impose another using \SetXYSeparator{<the new separator>}.  
Americans may prefer "," to "*" since they never use a 
comma as a decimal point; on the other hand, * may be more visible.

APPENDIX (I)  MERGING labelfig.tex LABELS INTO AN EPSF GRAPHICS OBJECT.

     As promised in the introduction, here is a recipe useful for
publishers. It works at least on Macintosh and at least for vectorized
graphics and Adobe type1 fonts.  (There is surely a similar recipe for
PCs under MSWindows.)

 (a)  Use boxedeps.tex utility to integrate the figure given by the eps
file, "x.eps" say, with a visible frame around it.  See
\ShowDisplacementBoxes command in boxedeps.tex.  To get precise results
automatically it is important to use the \Trim... commands of
boxedeps.tex making the "DisplacementBox" neatly fit the figure.

 (b)  Use the TeX printer driver and LaserWriter (versions >= 8.1.1) to
export to an EPSF the DVI page containing the integrated, labelled
figure. You now have an EPS file  "xx.eps"  that contains too much, and at
the wrong scale, and at wrong position.

 (c)  Convert the EPSF to an Adode Illustrator format EPSF using
the shareware utility called epsConvert by Sam Weiss
1993-- (currently $25).

 (d)  In Illustrator (or a compatible program), group the labels and the
"DisplacementBox"; copy them to the clipboard and paste them into "x.ps".
This step requires that all the label fonts be "visible to the Macintosh.

 (e)  Translate and scale the pasted group consisting of the labels plus
the "DisplacementBox" so as to make the "DisplacementBox" the bounding
box of (labelless) figure represented by "x.eps".  At this point the
labels will be correctly placed on the figure "x.eps".

 (f)  Ungroup and delete the "DisplacementBox".  The result is the
desired single EPS file, "x+.eps" say, It contains the original figure
plus its labels.  

     Using grouping and ungrouping appropriately in "x+.eps", a
publisher's staff can very efficiently improve label positions etc.

APPENDIX II)  SOME EXOTIC APPLICATIONS

     The grid of labelfig.tex is analogous to a light-table in
classical page makeup with wax or latex glue.  In principle, you
can use it to compose any page from its indivisible parts.  This
even has some of the artisanal charm of classical paste-up
provided you have a fast screen preview to make the process
"interactive".

     In practice labelfig.tex is a tool for nonstandard jobs.
Here are a few going beyond the labelling already discussed.

(I)  GRAPHICS INTEGRATION.

     This is accomplished by treating the imported graphics
objects as labels.  The underlying graphics object is then
typically an empty  \vbox to <dimension>{\vfill} in a TeX
\midinsert...\endinsert construction.  A label line
might be of the form

   (.1*.1) \\

The exact form of the special command varies from driver to
driver.  However, in the case of encapsulated PostScript graphics
(EPSF norm), by relying on boxedeps.tex, one can have the
following standard syntax (independant of driver  (see
boxedeps.doc for details.
  
  (.1*.1) \BoxedEPSF{MyFigure scaled <scale in mils>}\\

This may be slow since it requires TeX to read the PostScript
file to read bounding box using many complex macros.  So you
may want to try

  (.1*.1) \EPSFSpecial{MyFigure}{<scale in mils>}\\

which is fast and driver independant, but it squashes the
bounding box, normally to its lower left corner.

     Similarly for graphics of the Macintosh PICT norm ---
using BoxedArt.tex (same sources) in place of boxedeps.tex.

     This approach to integration is to be recommended when
one is assembling a composite graphics object.

 (II)  COMMUTATIVE DIAGRAM ENHANCEMENT

     Commutative diagrams or arrays of mathematical objects
connected by arrows of various sorts are common in mathematics.
The mathematical objects require the use of TeX.  Recently TeX
acquired a good collection of arrows of all slopes --- that of
LamSTeX --- plus pwerful macros to build the diagrams.

     However, even the LamSTeX collection is often
inadequate; it lacks for example double shafted arrows, dotted
arrows and curved arrows. Fortunately it is possible to produce
such arrows on an individual basis using sophisticated graphics
programs such as Illustrator and AldusFreehand (both serving
the EPSF norm) or using Metafont (with its public domain norm).
Since the creation of each new arrow is a work of love, you
probably want to limit the number of arrows by using LamSTeX
for most arrows. The 40K commutative diagram module of LamSTeX
has been adapted to work with AmSTeX and a copy may be posted
with LabelFig and related files. Unfortunately no one has yet
offered a version that works with Plain TeX or LaTeX.

       Suffice it here to say that when the exotic arrow has
been somehow imported into TeX, labelfig.tex treats it as a
label that one affixes to the commutative diagram.  Two other
steps will be treated in separate notes, namely the matter of
extracting the dimension specifications for the arrow and the
construction of the arrow --- for these steps are far from
unique and often depend intimately on your computer environment. 
Notes for the Macintosh-Textures-Illustrator combination are
found in the file ExoticArrows.doc.

 (III) NESTING 

Ingenuity pays off in exploiting labelfig.tex. One can
mix graphics and typography quite freely.  labelfig.tex is good
for freeform or overlapping arrangements, while boxedeps.tex (or
BoxedArt.tex) is best for regimented non-overlapping
arrangements --- and the two can be combined.

     The default behavior of labelfig.tex is not ideal 
for nesting objects, because to prevent trouble for beginners
the register for labels is globally cleared when \AffixLabels
concludes.  But there are switches available

      \LabelsGlobal      \LabelsLocal

which change this.  To understand this, extend the above test 
by something like:

 \LabelsLocal

 \SetLabels
    (.5*.5) AAA\\
 \endSetLabels

 {
 \SetLabels
    (.5*.5) ZZZ\\
 \endSetLabels
   \AffixLabels{\FirstQuadrant}
 }

   \AffixLabels{\FirstQuadrant}

     There are however potential pitfalls.  Neither
labelfig.tex nor boxedeps.tex has been tested under extreme
conditions. Problems may occur if their procedures are
indiscriminately nested. For boxedeps.tex (not labelfig.tex)
there is a precise cause for worry, namely many of its
variables are "global", which means that TeX braces will not
provide the protection one might expect.

COMMAND SUMMARY FOR labelfig.tex

  Here [...] means optional (one or zero)
       [...]* means any number of such constructs

  \SetLabels
    [[<P>](<X><Sep><Y>) <label> \\]*
  \endSetLabels
  \ShowGrid  
  \AffixLabels{<the figure>}

   --- <P> is tack position, one of eleven or empty
              order irrelevant

                   \L\T      \T      \R\T

                   \L\E      \E      \R\E

                     \L               \R

                   \L\B      \B      \R\B

   --- (<X><Sep><Y>) insertion point;
  <Sep> is separator, = * by default;
  \SetXYSeparator{<Sep>} changes it.
   <X> and <Y> are real numbers

  --- <label> a label to attach 

  --- <the figure> the figure to label 

  \GlobalLabels (default)     
  \LocalLabels  setting for nested constructs.

 \Grids makes ALL grids appear; \HideGrid then makes just next disappear.
 \noGrids returns to default.  The commands are always global.

 \GridLineWidth{<dimension>} adjusts width of grid lines. Default is very
small, to give "hairline" effect. If your grid lines are missing try
setting \GridLineWidth{1pt}.

 \Edges#1 globally changes the edge size of all grids to the numerical 
value #1, which must be 1, 10, 100, or 1000.  The default is 1.

VERSION HISTORY.
 --- Jan 1993: \Edges#1 and [??] option after \SetLabels
 --- July 1992: \Grids, \noGrids, \HideGrid;
       Gridlines become hairlines; \GridLineWidth{<dimension>}.
 --- Oct 1991, Jan 1992: \SetXYSeparator{<Sep>},  \LabelsGlobal,
       \LabelsLocal.
 --- July 1991: first release

Address for bugs and other feedback:

        Raymond S\'eroul
        IREM and Lab. de Typographie Informatise
        Univ. Rene Descartes
        Strasbourg

    Tel 33-88-41-63-45
    Email:  A18645@FRCCSC21.BITNET

        Laurent Siebenmann
        Mathematique, Bat. 425,
        Univ de Paris-Sud,
        91405-Orsay,
        France

    Tel 33-1-6941-7949; 
    Email: lcs@topo.math.u-psud.fr

\def\scalefig#1{\epsfxsize #1\textwidth}
\def\defeq{\stackrel{\Delta}{=}}

\newcommand {\Ebb}{{\mathbb{E}}}

\newcommand{\Kmsc}{{\mathscr{K}}}
\newcommand{\Amsc}{{\mathscr{A}}}
\newcommand{\Bmsc}{{\mathscr{B}}}

\newtheorem{theorem}{Theorem}
\newtheorem{lemma}{Lemma}

\newtheorem{corollary}{Corollary}

\newtheorem{condition}{Condition}

\newtheorem{problem}{Problem}

\title{\LARGE {On Beamformer Design  for
Multiuser MIMO Interference Channels}}

\author{
Juho Park, {\em Student~Member,~IEEE}, Youngchul
Sung$^\dagger$\thanks{$^\dagger$Corresponding author}, {\em
Senior~Member,~IEEE}, and\\ H. Vincent Poor, {\em Fellow,~IEEE}
\thanks{Juho Park and Youngchul Sung are with the Dept. of Electrical Engineering,  KAIST, Daejeon 305-701, South
Korea. E-mail:\{jhp@ and ysung@ee.\}kaist.ac.kr and  H. Vincent
Poor is with Dept. of Electrical Engineering, Princeton
University, Princeton, NJ 08544, E-mail:poor@princeton.edu. This
work was supported by the Korea Research Foundation Grant funded
by the Korean Government (KRF-2008-220-D00079).}
}

\begin{document}

\maketitle

\begin{abstract}
This paper considers several linear beamformer design paradigms
for multiuser time-invariant multiple-input multiple-output
interference channels. Notably,   interference alignment and
sum-rate based algorithms such as the maximum
signal-to-interference-plus noise (max-SINR) algorithm are
considered.  Optimal linear beamforming under interference
alignment consists of two layers; an inner precoder and decoder
(or receive filter) accomplish interference alignment to eliminate
inter-user interference, and an outer precoder and decoder
diagonalize the effective single-user channel resulting from the
interference alignment by the inner precoder and decoder.  The
relationship between this two-layer beamforming  and the max-SINR
algorithm is established at high signal-to-noise ratio. Also, the
optimality of the max-SINR algorithm within the class of linear
beamforming algorithms, and its local convergence with exponential
rate, are established at high signal-to-noise ratio.
\end{abstract}


\section{Introduction}

Since interference alignment was shown in
\cite{Cadambe&Jafar:08IT} to achieve the maximum number of degrees
of freedom (DoF) in $K$-user (possibly time-varying) interference
channels, this technique has attracted considerable attention as a
candidate for handling interference in multiuser wireless
environments.  With interference alignment, each user achieves
almost half of the capacity achievable without interference and
the total sum rate of the system is given by
\begin{equation}
C = \frac{K}{2}\log(\mbox{SNR}) + o(\log(\mbox{SNR})),
\end{equation}
where the $o(\log(\mbox{SNR}))$ term decays faster than
$\log(\mbox{SNR})$ as the signal-to-noise ratio (SNR) increases.
Interference alignment can be classified into two categories:
signal level (or scale) alignment \cite{Bresler&Parekh&Tse:10IT,
Cadambe&Jafar&Shamai:09IT,Sridharan&Jafarian&Viswanath&Jafar:08Globecom,
Sridharan&Jafarian&Vishwanath&Jafar:08Arxiv,He&Yener:09ITW,Etkin&Ordentlich:09Arxiv,
Motahari&Gharan&MadddahAli&Khandani:09Arxiv2} and signal space
alignment  \cite{Cadambe&Jafar:08IT, Gou&Jafar:08Arxiv,
Jafar&Shamai:08IT,Maddah-Ali&Motahari&Khandani:08IT,Weingarten&Shamai&Kramer:07ITA}.
(For a nice survey of the literature in this area, see
\cite{Yetis&Gou&Jafar&Kayran:10SP}.) While the alignment in signal
scale lends tractability to  DoF characterization, interference
alignment in signal space provides an attractive way to realize
interference alignment in practice. The signal space can be
generated in several ways, such as by concatenating time symbols
or frequency bins as in \cite{Cadambe&Jafar:08IT} when the channel
is varying over time or frequency, or by using multiple-input
multiple-output (MIMO) antenna techniques. Of these two
approaches, MIMO techniques seem to be more robust and attractive
for realistic slowly-fading wireless channels
\cite{Yetis&Gou&Jafar&Kayran:10SP}. While much work has been done
on the feasibility analysis and DoF characterization of
interference alignment, in this paper we focus on the algorithmic
aspect of interference alignment in signal space based on MIMO
antennas in time-invariant channels. Up to now, several algorithms
have been proposed to design interference-aligning beamforming
matrices at transmitters and receivers
 for the practical setting of time-invariant MIMO interference channels
\cite{Gomadam&Jafar:08ArxiV,Peters&Heath:09ICASSP,Yu&Sung:10SP,Kumar&Xue:10ISIT,Sung&Park&Lee&Lee:10WCOM}.
While these approaches design interference-aligning beamforming
matrices, others have proposed algorithms to maximize the sum rate
directly since interference alignment is optimal at high SNR and
optimal only in terms of DoF  even then, i.e., the
$o(\log(\mbox{SNR}))$ term still exists in the sum rate achieved
by interference alignment. Among these latter types of algorithms,
the max-SINR algorithm of Gomadam et al.
\cite{Gomadam&Jafar:08ArxiV} and the sum-rate gradient algorithm
of Sung et al. \cite{Sung&Park&Lee&Lee:10WCOM} are noticeable and
promising. Whereas the latter is based on the gradient descent of
the sum rate as a cost function, the former is based on the idea
of channel reciprocity and also on the individual stream approach
rather than  on the individual user approach aggregating multiple
streams of a single user. Although Gomadam et al. proposed the
max-SINR algorithm, its overall optimality and behavior were not
explored fully in \cite{Gomadam&Jafar:08ArxiV}.   In this paper,
we investigate the relationship among three beamformer design
algorithms: interference alignment, max-SINR and sum-rate gradient
algorithms, and show the optimality of the max-SINR algorithm
within the class of linear beamforming algorithms at high SNR.
Optimizing the beamforming filters for multiuser time-invariant
MIMO interference channels is not a simple problem as noted in
\cite{Gomadam&Jafar:08ArxiV}, and the analysis of such algorithms
as the max-SINR algorithm  is not a trivial task either due to the
nonconvex nature of the problem. Our approach to this analytical
task is based on  {\em fixed point analysis}
\cite{Censor&Zenios:book}. It is not difficult to show that
optimal linear beamforming under interference alignment consists
of two layers composed of inner and outer layers, which was
independently derived in \cite{Sung&Park&Lee&Lee:10WCOM}.  An
inner precoder and decoder\footnote{We will use the term 'linear
decoder' or simply 'decoder' for the receive filter in this
paper.} accomplish interference alignment to eliminate inter-user
interference, and an outer precoder and decoder diagonalize the
effective single-user channel resulting from the interference
alignment by the inner precoder and decoder. Based on  fixed point
analysis, we have shown the following properties  of the
considered algorithms and the relationship among them.
\begin{itemize}
\item[\textit{(i)}] At high SNR, all fixed points of the max-SINR algorithm with a
DoF guarantee are optimal two-layer beamformers as noted above. %
\item[\textit{(ii)}] The set of fixed points of the max-SINR algorithm includes the globally optimal linear beamformer at high SNR.%
\item[\textit{(iii)}] Any interference-aligning precoder is a fixed point of the sum-rate gradient algorithm regardless of the optimality of the outer precoder and decoder. %
\item[\textit{(iv)}] Finally, at high SNR, the max-SINR algorithm
converges exponentially to a fixed point when it is initialized
within a neighborhood of the fixed point.
\end{itemize}
Thus, the max-SINR algorithm is optimal within the class of linear
beamforming algorithms at high SNR in the sense of \textit{(ii)}.
Comparing \textit{(i)} and \textit{(iii)}, the max-SINR algorithm
has advantage over the sum-rate gradient algorithm in that it
yields not only interference alignment but also optimal outer
coders  at high SNR. This is because the max-SINR algorithm is
based on a stream-by-stream approach and this individual stream
approach adds  resolving power to the max-SINR algorithm compared
with the user-by-user algorithms.

This paper is organized as follows. The data model and background
are described in Section \ref{sec:systemmodel}. In Section
\ref{sec:optimaldesign}, we explain the two-layer linear precoder
and decoder structure for MIMO interference channels. In Section
\ref{sec:property}, we investigate the properties of the sum-rate
based beamformer design algorithms and the relationship with the
two-layer linear beamforming structure of Section
\ref{sec:optimaldesign}. In Section  \ref{sec:numerical} we
provide some numerical results, followed by conclusions in Section
\ref{sec:conclusion}.

\vspace{0.5em} \noindent  {\em Notation}

We will make use of standard notational conventions. Vectors and
matrices are written in boldface with matrices in capitals. All
vectors are column vectors.   For a matrix $\Abf$, $\Abf^T$ and
$\Abf^H$ indicate the transpose and Hermitian transpose of $\Abf$,
respectively,  and $\mbox{rank}(\Abf)$ represents the rank of
$\Abf$. $\Cc(\Abf)$ represents the column space of $\Abf$, i.e.,
the linear subspace spanned by the columns of $\Abf$. $|\Abf|$ and
$\Abf^\dagger$ denote the determinant and Moore-Penrose
pseudo-inverse of $\Abf$, respectively. For a linear subspace
$\Sc$, $\mbox{dim}(\Sc)$ denotes the dimension of $\Sc$, and
$\Sc^\perp$ represents the orthogonal complement of $\Sc$.
$\Ibf_n$ stands for the identity matrix of size $n$ (the subscript
is included only when necessary). For a vector $\abf$, $\| \abf
\|$ represents the 2-norm of $\abf$. The notation
$\xbf\sim\Nc(\mubf,\Sigmabf)$ means that the random vector $\xbf$
is complex Gaussian  with  mean vector $\mubf$ and  covariance
matrix $\Sigmabf$. $\Ebb\{\xbf\}$ denotes the expectation of
$\xbf$. ${\Amsc}\setminus{\Bmsc}$ denotes the relative complement
of a set ${\Amsc}$ in another set ${\Bmsc}$.

\section{Data Model and Background}
\label{sec:systemmodel}

\begin{figure}[t]
\centerline{
\begin{psfrags}
\psfrag{vd}[c]{{ $\vdots$}} %
\psfrag{d1}[l]{{ $d_1$}} %
\psfrag{d2}[l]{{ $d_2$}} %
\psfrag{dK}[l]{{ $d_K$}} %
\psfrag{V1}[c]{{ ${\bf V}_1$}} %
\psfrag{V2}[c]{{ ${\bf V}_2$}}  %
\psfrag{VK}[c]{{ ${\bf V}_K$}}  %
\psfrag{U1}[c]{{ ${\bf U}_1$}} %
\psfrag{U2}[c]{{ ${\bf U}_2$}} %
\psfrag{UK}[c]{{ ${\bf U}_K$}} %
\psfrag{N}[c]{{ $M$}} %
\psfrag{M}[c]{{ $M$}} %
\psfrag{H11}[c]{{ ${\bf H}_{11}$}} %
\psfrag{H21}[c]{{ ${\bf H}_{21}$}} %
\psfrag{HK1}[c]{{ ${\bf H}_{K1}$}} %
\psfrag{H12}[c]{{ ${\bf H}_{12}$}} %
\psfrag{H22}[c]{{ ${\bf H}_{22}$}} %
\psfrag{HK2}[c]{{ ${\bf H}_{K2}$}} %
\psfrag{H1K}[c]{{ ${\bf H}_{1K}$}} %
\psfrag{H2K}[c]{{ ${\bf H}_{2K}$}} %
\psfrag{HKK}[c]{{ ${\bf H}_{KK}$}} %
\scalefig{0.5}\epsfbox{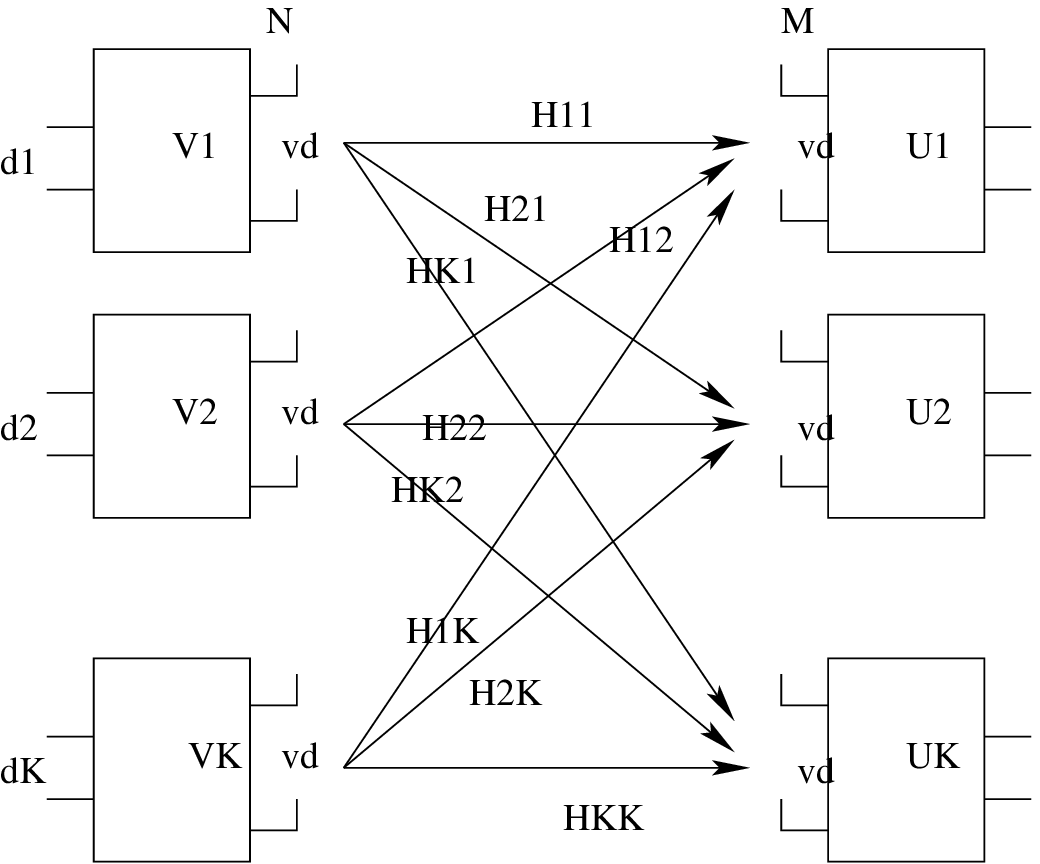}
\end{psfrags}
}\caption{$K$-user MIMO interference channel model.}
\label{fig:interf_ch}
\end{figure}

We consider a  $K$-user $M \times M$ MIMO interference channel in
which $K$ transmitters each having $M$ antennas simultaneously
transmit to $K$ receivers  each also having $M$ antennas, as shown
in Figure \ref{fig:interf_ch}. Due to interference each receiver
receives the desired signal from its corresponding transmitter and
also interference from other undesired transmitters. Thus, the
received signal vector at receiver $k$ at symbol time $t$ is given
by
\begin{eqnarray}
\ybf_k(t) &=& \Hbf_{kk}(t)\Vbf_{k}(t)\sbf_k(t)
  + \sum_{l \ne k}\Hbf_{kl}(t)\Vbf_l(t)\sbf_l(t) + \nbf_k(t), \label{eq:theVeryDataModel}\\
&=& \Hbf_{kk}(t)(\vbf_k^{(1)}(t) s_k^{(1)}(t) + \cdots
+\vbf_{k}^{(d_k)}(t)s_{k}^{(d_k)}(t))
  + \sum_{l \ne k}\Hbf_{kl}(t)\Vbf_{l}(t)\sbf_l(t) + \nbf_k(t), \nonumber
\end{eqnarray}
where ${\bf{H}}_{kl}(t)$ is an $M \times M$ channel matrix from
transmitter $l$ to receiver $k$ at symbol time $t$,
${\Vbf}_k(t)=[\vbf_k^{(1)}(t),\vbf_k^{(2)}(t),\cdots,\vbf_k^{(d_k)}(t)]$
is an $M \times d_k$ transmit beamforming matrix, ${\sbf}_k(t)$
$=[s_k^{(1)}(t),$ $\cdots,$ $s_k^{(d_k)}(t)]^T$ is a
$d_k$-dimensional transmit signal vector, $d_k$ is the number of
data streams for user $k$, and ${\bf{n}}_{k}(t)$ is an $M \times
1$ circularly symmetric complex Gaussian noise vector with
distribution $\Nc( {\mathbf{0}}, \Ibf_M)$.  The spatial signature
$\vbf_k^{(m)}(t)$ of the $m$-th stream of user $k$ has unit norm.
We assume that $s_k^{(m)}(t)$, $m=1,2,\cdots, d_k$, are
independent and ${\mathbb{E}}\{|s_k^{(m)}(t)|^2\}=P_k^{(m)}$ for
all $t=1,2,\cdots$, and the total transmit power of the overall
system is given by
\begin{equation} \label{eq:totalPowerConstraint}
P_t = \sum_{k=1}^{K}\sum_{m=1}^{d_k}P_k^{(m)}.
\end{equation}
We  assume that the  channels are time-invariant, i.e., the
channel matrices $\{\Hbf_{kl}(t)\}$ do not change over time. Thus,
we omit the time index $t$ from here on. We also assume that
channel information is known to all of the transmitters and the
receivers. Further, in this paper we consider the case of
$d=d_1=d_2=\cdots=d_K=M/2$ to allow for a maximum number of
degrees.

\subsection{Background}

In this subsection, we briefly recapitulate those results of the
previous works \cite{Gomadam&Jafar:08ArxiV} and
\cite{Sung&Park&Lee&Lee:10WCOM} that are relevant to the analysis
in the sequel.

\subsubsection{Interference alignment in signal space}

The basic idea of interference alignment via the signal space
approach \cite{Cadambe&Jafar:08IT} is to confine the interference
from the undesired transmitters within a linear subspace at the
receiver of dimension less than that of the received signal space
so that the remaining subspace can be used for interference-free
communication. Thus, the interference alignment condition is given
as follows
\cite{Gomadam&Jafar:08ArxiV,Yetis&Gou&Jafar&Kayran:10SP}.

\vspace{0.5em}
\begin{condition} \label{cond:perfectIA2}
There exist non-zero matrices $\{\Ubf_k:
\mbox{size}(\Ubf_k)=M\times d,~ \mbox{rank}(\Ubf_k) = d, ~~k
=1,2,\cdots, K \}$ and $\{\Vbf_l: \mbox{size}(\Vbf_l)=M\times
d,~\mbox{rank}(\Vbf_l) = d, ~~l=1,2,\cdots, K\}$ such that
\begin{eqnarray}
\Ubf_k^H{\bf{H}}_{kl}{\bf{V}}_l &=&{\mathbf{0}},~~  k \in \Kmsc
\defeq \{1,2,\cdots,K\},
~l\in \Kmsc \setminus \{k\}, \label{eq:IAcond2_1}\\
\mbox{and ~rank}(\Ubf_k^H \Hbf_{kk} \Vbf_k) &=& d, ~~ k \in \Kmsc.
\label{eq:IAcond2_2}
\end{eqnarray}
\end{condition}
\vspace{0.5em}

\noindent Here, $\Vbf_k$ and $ \Ubf_k =[\ubf_k^{(1)},\cdots,
\ubf_k^{(d)}]$ are the transmit beamforming matrix (or linear
precoder) and receive beamforming matrix (or linear decoder),
respectively, of user $k$. Thus, when interference alignment is
achieved by the transmit and receive beamforming matrices
$\{\Vbf_k\}$ and $\{ \Ubf_k\}$, $\Cc(\Ubf_k)$ is the orthogonal
complement of the aligned interference subspace generated by
$\sum_{l\ne k} \Hbf_{kl}\Vbf_l \sbf_l$ at user $k$, and we have
the following interference covariance matrix and its singular
value decomposition (SVD):
\begin{eqnarray}
\Zbf_k &\defeq& \sum_{l\ne k} \Hbf_{kl} \Vbf_l \mbox{diag}(P_l^{(1)},\cdots, P_l^{(d)}) \Vbf_l^H \Hbf_{kl}^H, \nonumber\\
&=&\Gammabf_k \mbox{diag}(\sigma_{k1},\cdots, \sigma_{k
d},0,\cdots,0) \Gammabf_k^H. \label{eq:JafarUV1}
\end{eqnarray}
Also, by Hermitian transposing (\ref{eq:IAcond2_1}) and summing
the terms over $k$, we have $\sum_{k\ne l} \Vbf_l^H \Hbf_{kl}^H
\Ubf_k$ $=$   $\Vbf_l^H\sum_{k\ne l}$ $( \Hbf_{kl}^H \Ubf_k) =0$
and
\begin{eqnarray}
\overleftarrow{\Zbf}_l &\defeq& \sum_{k\ne l} \overleftarrow{\Hbf}_{lk} \Ubf_k \mbox{diag}(\overleftarrow{P}_k^{(1)},\cdots, \overleftarrow{P}_k^{(d)})\Ubf_k^H \overleftarrow{\Hbf}_{lk}, \nonumber\\
&=&\overleftarrow{\Gammabf}_k
\mbox{diag}(\overleftarrow{\sigma}_{k1},\cdots,
\overleftarrow{\sigma}_{k d},0,\cdots,0)
\overleftarrow{\Gammabf}_k^H , \label{eq:JafarUV2}
\end{eqnarray}
where $\overleftarrow{\Hbf}_{lk}= \Hbf_{kl}^H$ and
$\overleftarrow{P}_k^{(m)}$ is the power associated with
$\ubf_k^{(m)}$. There exist efficient algorithms to obtain
interference aligning beamforming matrices, e.g.,
\cite{Gomadam&Jafar:08ArxiV, Peters&Heath:09ICASSP, Yu&Sung:10SP},
and thus interference-aligning beamforming matrices $\{\Ubf_k,
\Vbf_k\}$ can be acquired easily  using such algorithms.

\subsubsection{Sum rate and the related algorithms}
\label{subsubsec:SumRateAlgorithms} 

Although interference alignment achieves the maximum number of
degrees of freedom in multiuser MIMO interference channels, i.e.,
the rate achieved by interference alignment is within a constant
gap from the capacity regardless of the value of SNR at high SNR,
larger sum rate can be achieved by linear sum-rate maximizing
beamforming at low and intermediate SNR and even at high SNR.
Consider the $m$-th stream of user $k$ and the corresponding
received signal at receiver $k$ given by
\begin{equation} \label{eq:MIMO_IC}
\ybf_k = \Hbf_{kk} \vbf_k^{(m)} s_k^{(m)} + \sum_{j\ne m}
\Hbf_{kk} \vbf_k^{(j)} s_k^{(j)} + \sum_{l \ne
k}\Hbf_{kl}\Vbf_{l}\sbf_l + \nbf_k,
\end{equation}
where the covariance matrix of the overall interference and noise
for this stream is given by
\begin{equation}  \label{eq:sumrateExplainRbfkm}
\Rbf_k^{(m)} \defeq \sum_{l=1}^{K} \sum_{j=1}^{d}
  P_l^{(j)} \Hbf_{kl} \vbf_{l}^{(j)} \vbf_{l}^{(j)H} \Hbf_{kl}^H
- P_k^{(m)} \Hbf_{kk} \vbf_{k}^{(m)} \vbf_{k}^{(m)H} \Hbf_{kk}^H +
\Ibf.
\end{equation}
Thus, the signal-to-interference-plus-noise ratio (SINR)
maximizing receiver filter $\ubf_k^{(m)}$ for this stream is given
by the whitened matched filter \cite{Poor:book}:
\begin{equation} \label{eq:sumrateExplainOptubfkm}
{\ubf}_k^{(m)} = \frac{(\Rbf_k^{(m)})^{-1} \Hbf_{kk}
\vbf_{k}^{(m)}}
     {\|(\Rbf_k^{(m)})^{-1} \Hbf_{kk} \vbf_{k}^{(m)}\|}
\end{equation}
(the normalization for unit norm does not affect the SINR) and the
corresponding rate for the stream is given by
\begin{equation} \label{eq:individualRate}
C_k^{(m)}=\log \left ( 1+ P_k^{(m)} \vbf_k^{(m)H} \Hbf_{kk}^H
(\Rbf_k^{(m)})^{-1} \Hbf_{kk} \vbf_k^{(m)} \right).
\end{equation}
The overall sum rate of the system is given by
\begin{equation} \label{eq:sumrateIndividual}
C = \sum_{k=1}^{K} \sum_{m=1}^{d} C_k^{(m)}.
\end{equation}
Obtaining  sum-rate maximizing transmit beam vectors
$\{\vbf_k^{m}, k=1,\cdots,K, m=1,\cdots,d\}$ is not a simple
problem due to the nonconvex dependence structure of $C$ on
$\vbf_k^{(m)}$ via $(\Rbf_k^{(m)})^{-1}$ in
(\ref{eq:individualRate}). Thus, several researchers have proposed
iterative algorithms to design beamforming matrices for maximizing
the sum rate of the system directly, e.g.,
\cite{Gomadam&Jafar:08ArxiV,Sung&Park&Lee&Lee:10WCOM}. In
 \cite{Gomadam&Jafar:08ArxiV}, Gomadam et al. proposed the
{\em max-SINR algorithm} to design sum-rate maximizing linear
precoders and decoders based on the individual stream approach and
on channel reciprocity suggested from the duality between the
Gaussian multiple access channel (MAC)  and the Gaussian broadcast
channel (BC) \cite{Vishwanath&Jindal&Goldsmith:02ICC,
Jindal&Vishwanath&Goldsmith:04IT,Viswanath&Tse:03IT}. The max-SINR
algorithm based on the idea of reciprocity provides an effective
method to design sum-rate maximizing linear precoders and
decoders. The optimality and solution structure of this algorithm
will be discussed in later sections.

\vspace{0.5em} \noindent
 \textbf{The max-SINR algorithm (Gomadam et al. \cite{Gomadam&Jafar:08ArxiV})}
\vspace{0.5em}

\begin{enumerate}
[Step 1.] \item  Fix $\{P_k^{(m)}, k=1,\cdots,K, ~m=1,\cdots,d\}$
such that $P_t = \sum_{k=1}^{K}\sum_{m=1}^{d_k}P_k^{(m)}$, and initialize $n=0$ and  $\{\vbf_k^{(m)}[n]\}$. (Equal power allocation was used in \cite{Gomadam&Jafar:08ArxiV}.)%
\item (VU-step) Compute the receiver filters $\{\ubf_k^{(m)}[n]\}$
for all streams of all users from (\ref{eq:sumrateExplainRbfkm})
and (\ref{eq:sumrateExplainOptubfkm}). %
\item (UV-step) Exploiting channel reciprocity, compute
$\{\vbf_k^{(m)}[n+1]\}$ by  treating  $\{\ubf_k^{(m)}[n]\}$ from
Step 1 as the transmit vector and  changing the role of
transmitter and receiver. That is,
 obtain  the interference-plus-noise covariance matrix in
the reciprocal channel:
\[
\overleftarrow{\Rbf}_k^{(m)}[n] =  \sum_{l=1}^{K} \sum_{j=1}^{d}
  \overleftarrow{P}_l^{(j)} \overleftarrow{\Hbf}_{kl}
  \ubf_{l}^{(j)}[n]
  \ubf_{l}^{(j)}[n]^H \overleftarrow{\Hbf}_{kl}^H
- \overleftarrow{P}_k^{(m)} \overleftarrow{\Hbf}_{kk}
\ubf_{k}^{(m)}[n]
  \ubf_{k}^{(m)}[n]^H \overleftarrow{\Hbf}_{kk}^H + \Ibf,
\]
  and
\begin{equation}
{\vbf}_k^{(m)}[n+1] = \frac{(\overleftarrow{\Rbf}_k^{(m)}[n])^{-1}
\overleftarrow{\Hbf}_{kk} \ubf_{k}^{(m)}[n]}
     {\|(\overleftarrow{\Rbf}_k^{(m)}[n])^{-1} \overleftarrow{\Hbf}_{kk} \ubf_{k}^{(m)}[n]\|}.
\end{equation}
\item Increase $n$ and iterate Steps 2 and 3.
\end{enumerate}
\vspace{0.5em}

\noindent  Note that the max-SINR algorithm itself does not
consider interference alignment but tries to increase the sum rate
through designing better beamforming matrices to increase the
stream SINR.

On the other hand, Sung et al. proposed a different iterative
method to design sum-rate maximizing beamforming matrices based on
the gradient descent method and user-by-user approach in
\cite{Sung&Park&Lee&Lee:10WCOM}, which we call the sum-rate
gradient algorithm in this paper. Under the assumption of equal
power allocation, the rate for user $k$ based on linear
beamforming matrices $\{\Vbf_k\}$ for the model
(\ref{eq:theVeryDataModel}) is given by
\begin{eqnarray}
C_k
&=& \log \left|\Ibf+\frac{P_t}{Kd}(\Ibf+\Zbf_k)^{-1}\Hbf_{kk}\Vbf_k
    \Vbf_k^{H}\Hbf_{kk}^H \right|,  \label{eq:sumrategradCk} \\
&=& \log \left|(\Ibf+\Zbf_k)^{-1}\left(\Ibf+\Zbf_k+\frac{P_t}{Kd} \Hbf_{kk}\Vbf_k\Vbf_k^{H}\Hbf_{kk}^H\right) \right|, \\
&=& \log \left|(\Ibf+\Zbf_k)^{-1}\left(\Ibf+\frac{P_t}{Kd} \sum_{l=1}^{K}\Hbf_{kl}\Vbf_l\Vbf_l^{H}\Hbf_{kl}^H\right) \right|, \\
&=& \log \left|\Rbf_k\right| - \log\left|\Ibf+\Zbf_k\right|,
\end{eqnarray}
where the overall signal  covariance matrix $\Rbf_k$ for user $k$
is given by
\begin{equation}
\Rbf_k \defeq \sum_{l=1}^{K}  \frac{P_t}{Kd}  \Hbf_{kl} \Vbf_{l}
\Vbf_{l}^{H} \Hbf_{kl}^H + \Ibf,
\end{equation}
 and the overall
sum rate is given by
\begin{equation} \label{eq:sumrategradTotalRate}
C = \sum_{k=1}^K C_k.
\end{equation}
The direction of maximum increase of the functional
$C(\Vbf_1,\cdots,\Vbf_K)$ is given by the gradient of the
functional $C(\Vbf_1,\cdots,\Vbf_K)$ with respect to $\Vbf_k^o$,
where $\Vbf_k^o=(\Vbf_k^H)^T$, and it is obtained by the complex
gradient operator
\cite{Magnus&Neudecker:book,Brandwood:83IEEproc}:
\begin{equation}
\nabla_{\Vbf_k^o} C(\Vbf_1,\cdots,\Vbf_K)= \sum_{l=1}^{K}
\left(\frac{\partial\log|\Rbf_l|}{\partial\Vbf_k^o}
     -\frac{\partial\log|\Ibf+\Zbf_l|}{\partial\Vbf_k^o} \right).
\end{equation}
From the fact that for a matrix $\Cbf$
\begin{equation}
\frac{\partial\log|\Cbf(\{\Vbf_l\})|}{\partial\Vbf_k^o}
=\mathrm{tr}\left\{\Cbf(\{\Vbf_l\})^{-1}\frac{\partial
\Cbf(\{\Vbf_l\})}{\partial \Vbf_k^o}  \right\},
\end{equation}
the gradient of the sum rate  (\ref{eq:sumrategradTotalRate}) with
respect to $\Vbf_k^o$ is given by
\begin{equation}
\label{eq:gradient}
\nabla_{\Vbf_k^o} C(\Vbf_1,\cdots,\Vbf_K)=
 \sum_{l=1}^{K} \frac{P_t}{Kd}\Hbf_{lk}^H\Rbf_l^{-1}\Hbf_{lk}\Vbf_k
-\sum_{l\neq k}
\frac{P_t}{Kd}\Hbf_{lk}^H(\Ibf+\Zbf_l)^{-1}\Hbf_{lk}\Vbf_k.
\end{equation}
An algorithm was constructed in \cite{Sung&Park&Lee&Lee:10WCOM} to
update the  beam based on the gradient and to converge at least to
a local maximum. Note that the max-SINR algorithm is based on the
stream-by-stream approach (\ref{eq:MIMO_IC},
\ref{eq:individualRate}, \ref{eq:sumrateIndividual}) whereas the
sum rate gradient algorithm relies on the user-by-user approach
(\ref{eq:theVeryDataModel}, \ref{eq:sumrategradCk},
\ref{eq:sumrategradTotalRate}). The resulting difference between
the stream-by-stream  and user-by-user approaches will be
discussed in later sections.

\section{Two-Layer Linear Precoder and Decoder Structure\\ under Interference Alignment}
\label{sec:optimaldesign}

In this section, we consider optimal beamforming matrix design
under the interference alignment constraint, i.e.,
(\ref{eq:IAcond2_1}). As already noted in
\cite{Gomadam&Jafar:08ArxiV}, \cite{Yetis&Gou&Jafar&Kayran:10SP}
and \cite{Sung&Park&Lee&Lee:10WCOM}, once interference-aligning
beamforming matrices $\{\underbar{\Ubf}_k\}$ and
$\{\underbar{\Vbf}_k\}$ are given, any other matrices that
generate the same subspaces as $\{\underbar{\Ubf}_k\}$ and
$\{\underbar{\Vbf}_k\}$ are also interference-aligning. That is,
$\{\Vbf_k : \Vbf_k= \underbar{\Vbf}_k \Phibf_k, ~\Phibf_k \in
{\mathbb{C}}^{d \times d},  k\in\Kmsc\}$ and $\{\Ubf_k : \Ubf_k=
\underbar{\Ubf}_k \Thetabf_k, ~{\mathbf{\Theta}}_k \in
{\mathbb{C}}^{d \times d},  k\in\Kmsc \}$ are also interference
aligning beamforming matrices for the given channel for any
$\{\Phibf_k\}$ and $\{\Thetabf_k\}$ when $\{\underbar{\Ubf}_k\}$
and $\{\underbar{\Vbf}_k\}$ are interference-aligning. Thus, for a
given interference-aligning subspaces or matrices
$\{\underbar{\Ubf}_k\}$ and $\{\underbar{\Vbf}_k\}$, the matrices
$\{\Phibf_k\}$ and $\{\Thetabf_k\}$ can  be optimized further to
increase the overall sum rate.

\vspace{0.5em}
\begin{problem}[Optimization of linear precoder and decoder under
interference alignment] For given interference aligning subspaces
given by  the column spaces of $\{\underbar{\Vbf}_k\}$ and
$\{\underbar{\Ubf}_k\}$ composed of orthonormal
columns\footnote{The assumption of orthonormal columns does not
result in any loss.}, design optimal $\{\Phibf_k^*
\in\mathbb{C}^{d\times d}\}$ and $\{\Thetabf_k^*
\in\mathbb{C}^{d\times d}\}$ to maximize the sum rate.  Thus,
optimal interference-aligning linear precoder and decoder are
given by
\begin{equation}  \label{eq:optimalIAbeamProblem}
 \Vbf_k^{*}=\underline{\Vbf}_k\Phibf_k^* \quad \mbox{and} \quad
 \Ubf_k^{*}=\underline{\Ubf}_k\Thetabf_k^*.
\end{equation}
\end{problem}
\vspace{0.5em}

\noindent The difference between Problem 1 and the classical MIMO
beamforming problem is that in Problem 1 we have a restriction on
the choice of beamforming matrices within the class of
interference-aligning matrices, whereas the MIMO beamforming
problem has no such constraint. However, the solution to Problem 1
is  simple and here we present its solution for the purpose of
later discussion. (It was also derived in
\cite{Sung&Park&Lee&Lee:10WCOM}.) Let $\Pibf_{k} =
\underbar{\Ubf}_k$ be the first projection at receiver $k$. Then,
the projected signal at receiver $k$ is given by
\begin{eqnarray}
\rbf_k
&=& \Pibf_k^H \ybf_k, \label{eq:received_signal_1} \\
&=& \Pibf_k^H \Hbf_{kk} \Vbf_k \sbf_k
  + \Pibf_k^H \sum_{l\neq k} \Hbf_{kl}
    \Vbf_l \sbf_l + \Pibf_k \nbf_k, \label{eq:received_signal_2}  \\
&=& \underbar{\Ubf}_k^H \Hbf_{kk} \Vbf_{k} \sbf_k+ \tilde{\nbf}_k.
\label{eq:optimaldesign1}
 \end{eqnarray}
 The second term on the right-handed side (RHS) of
 (\ref{eq:received_signal_2}) disappears due to interference
 alignment.
 Substituting $\Vbf_k = \underbar{\Vbf}_k \Phibf_k$
into (\ref{eq:optimaldesign1}), we have
\begin{equation}
\rbf_k = \underbar{\Ubf}_k^H \Hbf_{kk}\underbar{\Vbf}_k \Phibf_k
\sbf_k + \tilde{\nbf}_k
\end{equation}
where $\tilde{\nbf}_k = \underbar{\Ubf}_k^H \nbf_k \sim
\Nc(0,\Ibf)$ since the columns of $\underbar{\Ubf}_k$ form an
orthonormal basis. Since $\underbar{\Ubf}_k$, $\underbar{\Vbf}_k$
and $\Hbf_{kk}$ are given, we define a $d \times d$ equivalent
channel matrix $\bar{\Hbf}_k
\defeq \underbar{\Ubf}_k^H\Hbf_{kk}\underbar{\Vbf}_k$ for user  $k$
(which is known to the transmitter and receiver). Let its SVD be
\begin{equation} \label{eq:compositechannelSVD}
\bar{\Hbf}_{k}=\bar{\Ubf}_{k} \bar{\Lambdabf}_{k}
\bar{\Vbf}_{k}^H.
\end{equation}
 Then, the equivalent single user channel for user $k$ under
interference alignment and the secondary projected signal are
given respectively by
\begin{eqnarray}
\rbf_k &=& \bar{\Hbf}_{k} \Phibf_k {\mathbf{s}}_k + \tilde{\nbf}_k
~~~\mbox{and}
\label{eq:optimalbeam11}\\
\rbf_k^ \prime &=& \Thetabf_k^H \rbf_k.
\end{eqnarray}
Now the channel model (\ref{eq:optimalbeam11}) is simply a
conventional single-user MIMO channel with  a known channel
$\bar{\Hbf}_{k}$ with indepedent noise $\tilde{\nbf}_k$; optimal
$\Phibf_k$ and $\Thetabf_k$ are given by the right and left
singular vectors of $\bar{\Hbf}_{k}$, respectively
\cite{Telatar:99}:
\begin{equation} \label{eq:compositechannelSVD1}
\Phibf_k^*= \bar{\Vbf}_{k} ~\mbox{and}~ \Thetabf_k^*=
\bar{\Ubf}_{k},
\end{equation}
and optimal power allocation $P_k^{(m)}$ is then performed by
water-filling across all\footnote{Power distribution across all
transmitters is a reasonable assumption with transmitter
collaboration in current wireless systems.} $Kd$ independent
parallel Gaussian channels under the total power constraint $P_t =
\sum_{k=1}^{K}\sum_{m=1}^{d_k}P_k^{(m)}$.  Thus, the best sum rate
achievable by interference alignment is given by
\begin{equation} \label{eq:maxRateByIA}
R_{IA}^* = \max_{\underbar{\Vbf}_k, \underbar{\Ubf}_k} R(
\underbar{\Vbf}_k \Phibf_k^* (\{\underbar{\Vbf}_k,
\underbar{\Ubf}_k\}), \underbar{\Ubf}_k \Thetabf_k^*
(\{\underbar{\Vbf}_k, \underbar{\Ubf}_k\})),
\end{equation}
where the dependence of $\Phibf_k^*$ and $\Thetabf_k^*$ on
$(\{\underbar{\Vbf}_k,$ $\underbar{\Ubf}_k\})$ is explicitly shown
and $R( \underbar{\Vbf}_k \Phibf_k^* (\{\underbar{\Vbf}_k,
\underbar{\Ubf}_k\}),$ $\underbar{\Ubf}_k \Thetabf_k^*
(\{\underbar{\Vbf}_k, \underbar{\Ubf}_k\}))$ is the best rate
achievable by the interference-aligning subspaces spanned by
$\{\underbar{\Vbf}_k, \underbar{\Ubf}_k\}$.

\begin{figure}[t]
\centerline{
\begin{psfrags}
\psfrag{vd}[c]{{ $\vdots$}} %
\psfrag{s11}[r]{{ $s_1^{(1)}$}} %
\psfrag{s1d}[r]{{ $s_1^{(d)}$}} %
\psfrag{s21}[r]{{ $s_2^{(1)}$}} %
\psfrag{s2d}[r]{{ $s_2^{(d)}$}} %
\psfrag{sk1}[r]{{ $s_k^{(1)}$}} %
\psfrag{skd}[r]{{ $s_k^{(d)}$}} %
\psfrag{d}[l]{{ $d$}} %
\psfrag{V1}[c]{{ $\underbar{\Vbf}_1$}} %
\psfrag{V2}[c]{{ $\underbar{\Vbf}_2$}}  %
\psfrag{VK}[c]{{ $\underbar{\Vbf}_K$}}  %
\psfrag{P1}[c]{{ ${\bf \Phi}_1$}} %
\psfrag{P2}[c]{{ ${\bf \Phi}_2$}}  %
\psfrag{P3}[c]{{ ${\bf \Phi}_K$}}  %
\psfrag{T1}[c]{{ ${\bf \Theta}_1$}} %
\psfrag{T2}[c]{{ ${\bf \Theta}_2$}}  %
\psfrag{T3}[c]{{ ${\bf \Theta}_K$}}  %
\psfrag{U1}[c]{{ $\underbar{\Ubf}_1$}} %
\psfrag{U2}[c]{{ $\underbar{\Ubf}_2$}} %
\psfrag{UK}[c]{{ $\underbar{\Ubf}_K$}} %
\psfrag{N}[c]{{ $M$}} %
\psfrag{M}[c]{{ $M$}} %
\psfrag{Pkm}[l]{{ $\Ebb\{|s_k^{(m)}|^2\} = P_k^{(m)}$}} %
\psfrag{inner}[c]{{inner precoder and decoder}} %
\psfrag{outer}[c]{{outer precoder and decoder}} %
\psfrag{Hbf}[c]{{ $\{\Hbf_{kl}\}$}} %
\scalefig{0.8}\epsfbox{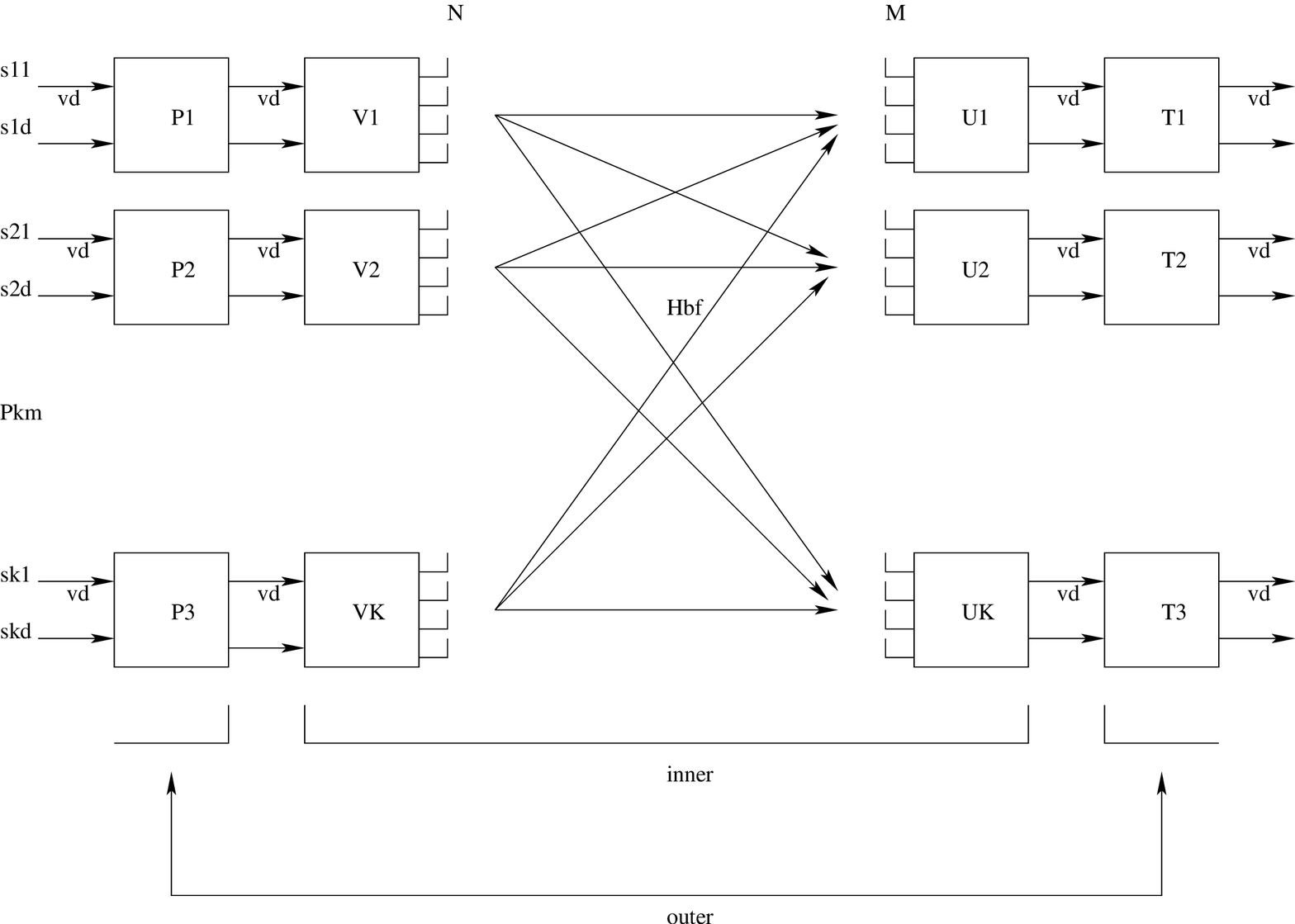}
\end{psfrags}
}\caption{Two-layer linear precoder and decoder structure under
interference alignment} \label{fig:inner_outer_precoder}
\end{figure}

 The optimal solution
(\ref{eq:optimalIAbeamProblem}) has an interesting precoding and
decoding structure, as shown in Fig.
\ref{fig:inner_outer_precoder}. That is, as noted previously,
optimal linear processing under interference alignment is composed
of two layers: inner precoders and decoders implement {\em
interference alignment} to yield an other-user-interference-free
single user channel for each transmit-receive pair, and outer
precoders and decoders perform single-user optimal {\em channel
diagonalization} for the  single user channel resulting from the
inner processing.

\section{Properties of the Sum-Rate Based Algorithms and their Relationship with the Two-Layer Beamformer Structure }
\label{sec:property}

In this section, we investigate the properties and solution
structure of the sum-rate-based beamformer design algorithms in
Section \ref{subsubsec:SumRateAlgorithms}, and establish an
optimality of the max-SINR algorithm and the relationship between
the sum-rate based algorithms and the two-layer linear beamformer
of the previous section at high SNR. Under an assumption of
sufficiently high SNR, we can use equal power distribution
$P_k^{(m)}=P_t/Kd$. We begin with an invariance property of the
max-SINR algorithm for interference-aligning subspace, given in
the following theorem.

\vspace{1em}
\begin{theorem} \label{theorem:maxSINRinvariance}
Any interference-aligning subspace is invariant under one
composite iteration (composed of one VU step and one UV step) of
the max-SINR algorithm at sufficiently high SNR. That is, when a
set of interference-aligning beamforming matrices
$\{\Vbf_k,\Ubf_k\}$ are input to the iteration of the max-SINR
algorithm, the resulting matrices after iteration have the same
subspace and thus maintains the interference alignment property at
sufficiently high SNR.
\end{theorem}

\vspace{1em}
\begin{proof}
Let any interference-aligning beamforming matrices $\Vbf_k
=[{\vbf}_k^{(1)},{\vbf}_k^{(2)},\cdots,{\vbf}_k^{(d)}]$ and
${\Ubf}_k =[{\ubf}_k^{(1)},{\ubf}_k^{(2)},\cdots,{\ubf}_k^{(d)}]$
be the input to the max-SINR algorithm at step $n$ (i.e.,
$\Vbf_k[n]={\Vbf}_k$ and $\Ubf_k[n-1]={\Ubf}_k$ for all $k$).  For
the $m$-th stream of user $k$, the receive filter is obtained by
the max-SINR algorithm as {\footnotesize
\begin{eqnarray}
{\mathbf{u}}_k^{(m)}[n] &=&
   {({\mathbf{R}}_k^{(m)}[n])^{-1} {\mathbf{H}}_{kk} {\mathbf{v}}_k^{(m)}[n]}
   \big/ {\|({\mathbf{R}}_k^{(m)}[n])^{-1} {\mathbf{H}}_{kk} {\mathbf{v}}_k^{(m)}[n]\|}, \nonumber \\
&=&
    \left(
    \sum_{l\neq k} \frac{P_t}{Kd} \Hbf_{kl}{\Vbf}_l {\Vbf}_l^{H} \Hbf_{kl}^{H}
   +\sum_{j\neq m} \frac{P_t}{Kd} \Hbf_{kk}{\vbf}_{k}^{(j)} {\vbf}_{k}^{(j)H} \Hbf_{kk}^H
   +\Ibf
\right)^{-1} \Hbf_{kk} {\vbf}_k^{(m)}
\bigg/ {\| \cdot \|}, \nonumber  \\
&\stackrel{(a)}{=}& \left( \left[ \qbf_{k,m}^{(1)}\
\qbf_{k,m}^{(2)}\ \cdots\ \qbf_{k,m}^{(M)} \right] \left[
\begin{array}{cccc}
\lambda_{k,m}^{(1)} + 1 &  \cdots & 0 & 0 \\
\vdots &    \ddots              &  0  & 0 \\
0 & 0 & \lambda_{k,m}^{(M-1)}+1  & 0 \\
0 & 0 &      0           & 1 \\
\end{array} \right]
\left[ \begin{array}{c}
\qbf_{k,m}^{(1)H} \\
\qbf_{k,m}^{(2)H} \\
\vdots                      \\
\qbf_{k,m}^{(M)H}
\end{array} \right]
\right)^{-1}
\Hbf_{kk} {\vbf}_k^{(m)} \big/ \| \cdot \|, \label{eq:theorem4_svd_q}  \\
&\stackrel{(b)}{=}&
\qbf_{k,m}^{(M)}\qbf_{k,m}^{(M)H} \Hbf_{kk} {\vbf}_k^{(m)} \bigg/ \|\cdot\|, ~~~\mbox{as $P_t \rightarrow \infty$, i.e., for sufficiently high SNR}, \nonumber \\
&=&
\qbf_{k,m}^{(M)} r_{k,m} e^{j\omega_{k,m}} \big/ \|\cdot\|~~~\mbox{for  a scalar  $r_{k,m}e^{j\omega_{k,m}}=\qbf_{k,m}^{(M)H}\Hbf_{kk}{\vbf}_k^{(m)}$}, \label{eq:thm4_phase1} \\
&=& \qbf_{k,m}^{(M)} e^{j\omega_{k,m}}. \nonumber
\end{eqnarray}
} Equality (a) is because $\{{\Vbf}_k\}$ satisfy the interference
alignment condition, resulting in $\mbox{rank}(\sum_{l\neq k}$
$\Hbf_{kl}{\Vbf}_l {\Vbf}_l^{H} \Hbf_{kl}^{H}) = d=M/2$ (see
(\ref{eq:JafarUV1})), and the covariance matrix of the additional
inter-stream interference has rank $d-1$. Thus, the total rank is
$M-1$. Equality (b) is by the assumption of sufficiently high SNR.
Collecting all $m=1,2,\cdots,d$, the VU step of the max-SINR
algorithm yields the receive beamforming matrix for user $k$ as
\begin{equation}
\Ubf_k[n]
=[\qbf_{k,1}^{(M)}e^{j\omega_{k,1}},\ \qbf_{k,2}^{(M)}e^{j\omega_{k,2}},\ \cdots,
  \qbf_{k,d}^{(M)}e^{j\omega_{k,d}}]. \label{eq:maxSINRUoutdn}
\end{equation}
With this $\Ubf_k[n]$ as the input to the UV step of the
iteration, we obtain a new transmit beamforming matrix
$\Vbf_{k}[n+1]$. For the $m$-th stream of user $k$, {\scriptsize
\begin{eqnarray}
{\mathbf{v}}_k^{(m)}[n+1] &=&
{(\overleftarrow{{\mathbf{R}}}_k^{(m)}[n])^{-1}\overleftarrow{{\mathbf{H}}}_{kk}
    {\mathbf{u}}_k^{(m)}[n]}
    \big/ {\|(\overleftarrow{{\mathbf{R}}}_k^{(m)}[n])^{-1}\overleftarrow{{\mathbf{H}}}_{kk}
    {\mathbf{u}}_k^{(m)}[n]\|} \nonumber \\
&=& \left( \sum_{l\neq k} \frac{P_t}{Kd}
\overleftarrow{\Hbf}_{kl}\Ubf_l[n] \Ubf_l[n]^H
\overleftarrow{\Hbf}_{kl}^{H} + \sum_{j\neq m} \frac{P_t}{Kd}
\overleftarrow{\Hbf}_{kk} \ubf_{k}^{(j)}[n] \ubf_{k}^{(j)}[n]^H
\overleftarrow{\Hbf}_{kk}^H + \Ibf \right)^{-1}
\overleftarrow{\Hbf}_{kk} \ubf_k^{(m)}[n]
\bigg/ {\| \cdot \|} \nonumber \\
&\stackrel{(a)}{=}& \left( \left[ \wbf_{k,m}^{(1)}\
\wbf_{k,m}^{(2)}\ \cdots\ \wbf_{k,m}^{(M)} \right] \left[
\begin{array}{cccc}
\sigma_{k,m}^{(1)} + 1 &  \cdots & 0 & 0 \\
\vdots &    \ddots              &  0  & 0 \\
0 & 0 & \sigma_{k,m}^{(M-1)}+1  & 0 \\
0 & 0 &      0           & 1 \\
\end{array} \right]
\left[ \begin{array}{c}
\wbf_{k,m}^{(1)H} \\
\wbf_{k,m}^{(2)H} \\
\vdots \\
\wbf_{k,m}^{(M)H}
\end{array} \right]
\right)^{-1}
\overleftarrow{\Hbf}_{kk}\ubf_k^{(m)}[n] \big/ \| \cdot \| \label{eq:theorem4_svd_w}\\
&=& \wbf_{k,m}^{(M)}\wbf_{k,m}^{(M)H}
\overleftarrow{\Hbf}_{kk}\ubf_k^{(M)}[n]
\bigg/ \|\cdot\|, ~~\mbox{as $P_t \rightarrow \infty$, i.e., for sufficiently high SNR,}~  \nonumber \\
&=& \wbf_{k,m}^{(M)} r^\prime_{k,m}e^{j\varphi_{k,m}} \big/
\|\cdot\| ~~~\mbox{for  scalar  $r^\prime_{k,m} e^{j\varphi_{k,m}}
=\wbf_{k,m}^{(M)H}\overleftarrow{\Hbf}_{kk}\ubf_k^{(m)}[n]$} \label{eq:theorem4scalarwbf}  \\
&=& \wbf_{k,m}^{(M)} e^{j\varphi_{k,m}},\nonumber
\end{eqnarray}
} where equality (a) is by Lemma \ref{lemma:UbfkTheSame}, i.e.,
$\Cc(\Ubf_k[n]) = \Cc({\Ubf}_k)$ ($\Ubf_k[n]$ is also
interference-aligning); thus,
$\mbox{rank}(\overleftarrow{{\Zbf}}_k)=d=M/2$ and
$\overleftarrow{{\mathbf{R}}}_k^{(m)}[n]$ has rank $M-1$ in total,
where
\begin{equation}
 \overleftarrow{{\Zbf}}_k= \frac{P_t}{Kd} \sum_{l\neq k}
\overleftarrow{\Hbf}_{kl} \Ubf_l[n] \Ubf_l[n]^{H}
\overleftarrow{\Hbf}_{kl}^H.
\end{equation}
 The filters for other streams can be obtained similarly.
Combining the filters for all streams of user $k$, we have
\begin{equation}
\Vbf_k[n+1]=[\wbf_{k,1}^{(M)}e^{j\varphi_{k,1}},\ \wbf_{k,2}^{(M)}e^{j\varphi_{k,2}},\
\cdots,\ \wbf_{k,d}^{(M)}e^{j\varphi_{k,d}}].
\label{eq:theorem4Vbfknp1}
\end{equation}
By similar argument as in Lemma \ref{lemma:UbfkTheSame}, we have
\begin{equation}  \label{eq:theorem4Vbfknp1subspace}
 \Cc(\Vbf_k[n+1])
=\Cc([\wbf_{k,1}^{(M)},\ \cdots, \wbf_{k,d}^{(M)}])
=\Cc(\overleftarrow{{\Zbf}}_k)^\bot =\Cc({\Vbf}_k).
\end{equation}
Thus, we have
\begin{equation}
\Cc(\Vbf_k[n+1])=\Cc({\Vbf}_k)=\Cc(\Vbf_k[n]).
\end{equation}
That is, one composite iteration of the max-SINR algorithm
preserves an interference-aligning subspace at sufficiently high
SNR.
\end{proof}

\vspace{1em}
\begin{lemma} \label{lemma:UbfkTheSame}
For the $\Ubf_k[n]$ in (\ref{eq:maxSINRUoutdn}) obtained by the VU
step of the max-SINR iteration, the following holds:
\begin{equation}       \label{eq:lemma1FirstEq}
\Cc(\Ubf_k[n])={\mathcal{C}}([\qbf_{k,1}^{(M)},\ \qbf_{k,2}^{(M)},
\ \cdots, \qbf_{k,d}^{(M)}]) \subset {\mathcal{C}}({\Zbf}_k)^\bot
= {\mathcal{C}}({\Ubf}_k)
\end{equation}
where ${\Zbf}_k = \frac{P_t}{Kd}\sum_{l\neq k} \Hbf_{kl} {\Vbf}_l
{\Vbf}_l^{H}\Hbf_{kl}^H$. When the rank of $\Ubf_k[n]$ is $d$,
\begin{equation}   \label{eq:lemma1SecondEq}
{\mathcal{C}}(\Ubf_k[n]) = {\mathcal{C}}([\qbf_{k,1}^{(M)},\
\qbf_{k,2}^{(M)},\ \cdots,\ \qbf_{k,d}^{(M)}]) =
{\mathcal{C}}({\Zbf}_k)^\bot = {\mathcal{C}}({\Ubf}_k).
\end{equation}
\end{lemma}

\vspace{1em}
\begin{proof}
 Let the SVD of ${\Zbf}_k$ be
\begin{equation}
{\Zbf}_k=
\left[
\zbf_{k}^{(1)}\ \zbf_k^{(2)}\ \cdots \zbf_{k}^{(M)}
\right]
\left[ \begin{array}{lc}
\Sigmabf_k              & {\mathbf{0}}_{d\times d} \\
{\mathbf{0}}_{d\times d}  & {\mathbf{0}}_{d\times d}
\end{array} \right]
\left[ \begin{array}{c}
\zbf_{k}^{(1)H} \\
\zbf_{k}^{(2)H} \\
\vdots \\
\zbf_{k}^{(M)H}
\end{array} \right] \label{eq:SVD_Zk},
\end{equation}
where $\Sigmabf_k$ is a $d\times d$ diagonal matrix containing
non-zero singular values of ${\Zbf}_k$. (This is because
$\{{\Vbf}_k\}$ is interference-aligning.) Hence,
\begin{equation}    \label{eq:Lemma1ZbfperpUbfk}
\Cc({\Zbf}_k)^\bot
=\Cc([\zbf_k^{(d+1)},\cdots,\zbf_k^{(M)}])
=\Cc(\Ubf_k).
\end{equation}
Furthermore, we can see from (\ref{eq:theorem4_svd_q}) that
\begin{equation}
\qbf_{k,m}^{(M)} =
\Cc([\qbf_{k,m}^{(1)},\qbf_{k,m}^{(2)},\cdots,\qbf_{k,m}^{(M-1)}])^\bot
\quad \mbox{for all~ } m, \label{eq:lemma2q1}
\end{equation}
and  for all $m$
\begin{equation}
{\mathcal{C}}({\Zbf}_k)\subset {\mathcal{C}}({\Rbf}_k^{(m)}[n])=
 {\mathcal{C}}([\qbf_{k,m}^{(1)},\qbf_{k,m}^{(2)},\cdots,\qbf_{k,m}^{(M-1)}])
\label{eq:lemma2q2}
\end{equation}
since $\Zbf_k$ is contained in $\Rbf_k^{(m)}$.  Now,
(\ref{eq:lemma2q1}) and (\ref{eq:lemma2q2}) imply
\begin{equation}
\qbf_{k,m}^{(M)}={\mathcal{C}}([\qbf_{k,m}^{(1)},\cdots,\qbf_{k,m}^{(M-1)}])^\bot
\subset{\mathcal{C}}({\Zbf}_k)^\bot, \label{eq:lemma2q3}
\end{equation}
since ${\mathcal{C}}(\Abf)\subset{\mathcal{C}}(\Bbf)$
$\Rightarrow$
${\mathcal{C}}(\Bbf)^\bot\subset{\mathcal{C}}(\Abf)^\bot$. Since
$\qbf_{k,m}^{(M)} \in {\mathcal{C}}({\Zbf}_k)^\bot$ for each
$m\in\{1,2,\cdots,d\}$, we have
\begin{equation}\label{eq:lemma1middle}
{\mathcal{C}}([\qbf_{k,1}^{(M)},\qbf_{k,2}^{(M)},\cdots,\qbf_{k,d}^{(M)}])
\subset {\mathcal{C}}({\Zbf}_k)^\bot,
\end{equation}
and (\ref{eq:lemma1FirstEq}) follows from
(\ref{eq:Lemma1ZbfperpUbfk}) and (\ref{eq:lemma1middle}). When
$\mbox{rank}([\qbf_{k,m}^{(1)},\qbf_{k,m}^{(2)},\cdots,\qbf_{k,m}^{(d)}])=d=
\mbox{rank}({\Zbf}_k)$, two subspaces are the same, i.e.,
\begin{equation}
{\mathcal{C}}([\qbf_{k,1}^{(M)},\qbf_{k,2}^{(M)},\cdots,\qbf_{k,d}^{(M)}])
={\mathcal{C}}({\Zbf}_k)^\bot = \Cc(\Ubf_k).
\end{equation}
\end{proof}

\vspace{0.5em} \noindent Next, we investigate the fixed point
structure of the max-SINR algorithm by showing the relationship
between the (coder) optimal two-layer precoder and decoder
structure in Section \ref{sec:optimaldesign} and the max-SINR
algorithm.

\vspace{0.5em}
\begin{theorem} \label{theo:TwoLayerFixedPointOfmaxSINR}
The  two-layer linear beamforming solution
(\ref{eq:optimalIAbeamProblem}) with optimal outer coder under
interference alignment, i.e., $\{\Vbf_k^* = \underbar{\Vbf}_k
\Phibf_k^*(\{\underbar{\Vbf}_k,\underbar{\Ubf}_k\}), \Ubf_k^* =
\underbar{\Ubf}_k
\Thetabf_k^*(\{\underbar{\Vbf}_k,\underbar{\Ubf}_k\})$, is a fixed
point of the max-SINR algorithm for any interference-aligning
matrices $\{\underbar{\Vbf}_k,\underbar{\Ubf}_k\}$ at sufficiently
high SNR.
\end{theorem}

\vspace{1em}
\begin{proof}
Set  $\{\Vbf_k^* = \underbar{\Vbf}_k \Phibf_k^*\}$  as the input
to the VU step of the max-SINR iteration. Let
$\Ubf_k[n]=[\qbf_{k,1}^{(M)}e^{j\omega_{k,1}}, \cdots,$
$\qbf_{k,d}^{(M)}e^{j\omega_{k,d}}]$ of (\ref{eq:maxSINRUoutdn})
be the corresponding output of the VU step (the notation here
follows Theorem \ref{theorem:maxSINRinvariance}). From
(\ref{eq:theorem4_svd_q}), ~$\qbf_{k,m}^{(M)}\  \bot\
{\mathcal{C}}([\qbf_{k,m}^{(1)}, \qbf_{k,m}^{(2)}, \cdots,
\qbf_{k,m}^{(M-1)}])=\Cc(\Rbf_k^{(m)})$ and this implies
\begin{eqnarray}
0 &=& \qbf_{k,m}^{(M)H} \left(\sum_{l\neq
k}\Hbf_{kl}\Vbf^*_{l}\Vbf^{*H}_l\Hbf_{kl}^H
     +\sum_{j\neq m}\Hbf_{kk}\vbf_k^{*(j)}\vbf_k^{*(j)H}\Hbf_{kk}^H \right)\qbf_{k,m}^{(M)} \\
  &=& \qbf_{k,m}^{(M)H}
\left(\sum_{l\neq
k}\Hbf_{kl}\Vbf^*_{l}\Vbf^{*H}_l\Hbf_{kl}^H\right)\qbf_{k,m}^{(M)}
     +\qbf_{k,m}^{(M)H}\left(\sum_{j\neq m}\Hbf_{kk}\vbf_k^{*(j)}\vbf_k^{*(j)H}\Hbf_{kk}^H\right)\qbf_{k,m}^{(M)}. \label{eq:theoremd=nfixedP1}
\end{eqnarray}
From Lemma \ref{lemma:UbfkTheSame} we have
\begin{equation} \label{eq:theoremd=nfixedP2}
{\mathcal{C}}([\qbf_{k,1}^{(M)},\ \qbf_{k,2}^{(M)},\ \cdots,\
\qbf_{k,d}^{(M)}]) \subset {\mathcal{C}}(\Ubf_k^*) =
{\mathcal{C}}(\underbar{\Ubf}_k).
\end{equation}
Because of this, the first term on the RHS in
(\ref{eq:theoremd=nfixedP1}) is nulled out by the interference
alignment condition. Also, (\ref{eq:theoremd=nfixedP2}) implies
that for each $m$
\begin{equation} \label{eq:Theorem2proofqbfkmM}
\qbf_{k,m}^{(M)} = \underbar{\Ubf}_k \xbf_{km}
\end{equation}
for some vector $\xbf_{km}$. Denote $\Vbf_k^*$ as
\begin{equation} \label{eq:Theorem2proofVbfk}
\Vbf_k^* =[\vbf_{k}^{*(1)}, ~\vbf_{k}^{*(2)}, \cdots,
~\vbf_{k}^{*(d)}] = \underbar{\Vbf}_k \Phibf_k^* =
\underbar{\Vbf}_k [\phibf_k^{*(1)},\phibf_k^{*(2)},
\cdots,\phibf_k^{*(d)}],
\end{equation}
where  $\Phibf_k^*$ is a unitary matrix composed of the right
singular vectors of the single-user equivalent channel
$\bar{\Hbf}_{k} =
\underline{\Ubf}_k^H\Hbf_{kk}\underline{\Vbf}_{k}$ in
(\ref{eq:compositechannelSVD}) with  SVD $
\bar{\Hbf}_{k}=\bar{\Ubf}_{k} \bar{\Lambdabf}_{k}\bar{\Vbf}_{k}^H
= \Thetabf_k^*\bar{\Lambdabf}_{k}\Phibf^{*H}_k$ (see
(\ref{eq:compositechannelSVD1})). Since both terms of the RHS of
(\ref{eq:theoremd=nfixedP1}) are nonnegative, we have for the
second term
\begin{eqnarray}
0
&=&\textstyle \qbf_{k,m}^{(M)H}(\sum_{j\neq m} \Hbf_{kk}\vbf_{k}^{*(j)}\vbf_{k}^{*(j)H}\Hbf_{kk}^H )\qbf_{k,m}^{(M)}, \nonumber  \\
&=&\textstyle \sum_{j\neq m} \qbf_{k,m}^{(M)H}\Hbf_{kk}\vbf_{k}^{*(j)}\vbf_{k}^{*(j)H}\Hbf_{kk}^H\qbf_{k,m}^{(M)},  \nonumber \\
&=&\textstyle \sum_{j\neq m} \left|\qbf_{k,m}^{(M)H}\Hbf_{kk}\vbf_{k}^{*(j)}\right|^2,  \nonumber \\
&\stackrel{(a)}{=}&\textstyle \sum_{j\neq m} \left|\xbf_{km}^H \underline{\Ubf}_k^H\Hbf_{kk}\underline{\Vbf}_{k}\phibf_{k}^{*(j)} \right|^2,  \nonumber \\
&\stackrel{(b)}{=}&\textstyle \sum_{j\neq m} \left|\xbf_{km}^H \bar{\Hbf}_{k} \phibf_{k}^{*(j)} \right|^2,  \nonumber \\
&\stackrel{(c)}{=}&\textstyle \sum_{j\neq m} \left|\xbf_{km}^H (\Thetabf_{k}^* \bar{\Lambdabf}_{k}\Phibf^{*H}_k) \phibf_{k}^{*(j)} \right|^2,  \nonumber \\
&=&\textstyle \sum_{j\neq m} \left|\xbf_{km}^H (\sum_{i=1}^{d} \bar{\lambda}_{k}^{(i)}\thetabf_k^{*(i)}\phibf_{k}^{*(i)H}) \phibf_{k}^{*(j)} \right|^2,  \nonumber \\
&\stackrel{(d)}{=}&\textstyle \sum_{j\neq
m}\left|\xbf_{km}^H(\bar{\lambda}_{k}^{(j)}\thetabf_k^{*(j)}) \right|^2, \nonumber \\
&=&\textstyle \sum_{j\neq m} \left|\bar{\lambda}_{k}^{(j)}
\xbf_{km}^H \thetabf_k^{*(j)} \right|^2,
\label{eq:theorem5pf2termzero1}
\end{eqnarray}
where (a) is by (\ref{eq:Theorem2proofqbfkmM}) and
(\ref{eq:Theorem2proofVbfk}), (b) is by the definition of the
equivalent channel,  (c) is by its SVD and (d) is because
$\Phibf_k^*$ is unitary. Since each term in the summation in
(\ref{eq:theorem5pf2termzero1}) is non-negative, each term (i.e.,
for each $j$) is zero for the sum to be zero. Therefore, the
unique $\xbf_{km}$ satisfying this is given by
\begin{equation} \label{eq:theorem5fixedPxbf}
\xbf_{km}=\thetabf_k^{*(m)}
\end{equation}
since $\Thetabf_k^*$ is unitary, and thus
\begin{equation} \label{eq:theorem5qkiopt}
 \qbf_{k,m}^{(M)}
 =\underline{\Ubf}_k \thetabf_k^{*(m)}.
\end{equation}
The filters for other streams can be obtained similarly. Combining
all the streams yields
\begin{eqnarray}
\Ubf_k[n]
&=&[\qbf_{k,1}^{(M)}e^{j\omega_{k,1}}, \
\qbf_{k,2}^{(M)}e^{j\omega_{k,2}}, \ \cdots,\
\qbf_{k,d}^{(M)}e^{j\omega_{k,d}}],\\
&=&[\underline{\Ubf}_k\thetabf_k^{*(1)}e^{j\omega_{k,1}},\
\underline{\Ubf}_k\thetabf_k^{*(2)}e^{j\omega_{k,2}},\ \cdots,\
\underline{\Ubf}_k\thetabf_k^{*(d)}e^{j\omega_{k,d}}],\\
&=&\underline{\Ubf}_k
\Thetabf_k^* \mbox{diag}(e^{j\omega_{k,1}},e^{j\omega_{k,2}}, \cdots, e^{j\omega_{k,d}}),\\
&=& [\ubf_k^{*(1)}e^{j\omega_{k,1}},\
\ubf_k^{*(2)}e^{j\omega_{k,2}},\
\cdots, ~\ubf_k^{*(d)}e^{j\omega_{k,d}}].
\end{eqnarray}
This is valid for  all  $k$  (i.e., for all users).

Now consider the UV step of the iteration from
$\Ubf_k[n]=\Ubf_k^{*}\mbox{diag}(e^{j\omega_{k,1}},\cdots,
e^{j\omega_{k,d}})$ to $\Vbf_k[n+1]$. From
(\ref{eq:theorem4Vbfknp1}) in the proof of Theorem
\ref{theorem:maxSINRinvariance}, we have
\[
\Vbf_k[n+1]=[\wbf_{k,1}^{(M)}e^{j\varphi_{k,1}},\
\wbf_{k,2}^{(M)}e^{j\varphi_{k,2}},\ \cdots,\
\wbf_{k,d}^{(M)}e^{j\varphi_{k,d}}] \] and
\eqref{eq:theorem4_svd_w} shows that $\wbf_{k,m}^{(M)}\ \bot\
{\mathcal{C}}([\wbf_{k,m}^{(1)},\wbf_{k,m}^{(2)},\cdots,\wbf_{k,m}^{(M-1)}])=\Cc(\overleftarrow{\Rbf}_k^{(m)})$.
From this and (\ref{eq:theorem4_svd_w}) we have
\begin{eqnarray}
0 &=& \wbf_{k,m}^{(M)H} \left(\sum_{l\neq
k}\overleftarrow{\Hbf}_{kl}{\Ubf}_{l}[n]
     {\Ubf}_l^H[n]\overleftarrow{\Hbf}_{kl}^H
     +\sum_{j\neq m}\overleftarrow{\Hbf}_{kk}\ubf_k^{(j)}[n](\ubf_k^{(j)}[n])^H\overleftarrow{\Hbf}_{kk}^H
\right)\wbf_{k,m}^{(M)} \\
&=& \wbf_{k,m}^{(M)H} \left(\sum_{l\neq
k}\overleftarrow{\Hbf}_{kl}{\Ubf}^*_{l}
    {\Ubf}^{*H}_l\overleftarrow{\Hbf}_{kl}^H \right)\wbf_{k,m}^{(M)}
   +\wbf_{k,m}^{(M)H}
\left(\sum_{j\neq
m}\overleftarrow{\Hbf}_{kk}\ubf_k^{*(j)}\ubf_k^{*(j)H}\
    \overleftarrow{\Hbf}_{kk}^H \right)\wbf_{k,m}^{(M)}.
    \label{eq:theorem5fixedP11}
\end{eqnarray}
By (\ref{eq:theorem4Vbfknp1subspace}) we have
\begin{equation}
\Cc([\wbf_{k,1}^{(M)},\wbf_{k,2}^{(M)}, \cdots,\wbf_{k,d}^{(M)}])
\subset \Cc (\Vbf_k^*) = \Cc(\overleftarrow{\Zbf}_k^*)^\perp,
\end{equation}
where $\overleftarrow{\Zbf}_k^* =(P_t/Kd)\sum_{l\neq k}
\overleftarrow{\Hbf}_{kl}{\Ubf}_{l}^*{\Ubf}_l^{*H}\overleftarrow{\Hbf}_{kl}^H$.
This implies that the first term on the RHS of
(\ref{eq:theorem5fixedP11}) is zero and for each $m$
\[
\wbf_{k,m}^{(M)}=\underline{\Vbf}_k\ybf_{km}
\]
for some vector $\ybf_{km}$. By a procedure similar to that used
to obtain (\ref{eq:theorem5fixedPxbf}) we obtain the unique
$\ybf_{km}=\phibf_k^{*(m)}$. Hence,
$\wbf_{k,m}^{(M)}=\underline{\Vbf}_k\phibf_k^{*(m)}$.  Combining
all the streams yields,
\begin{eqnarray}\label{eq:theorem5conclusionwithrotation}
\Vbf_k[n+1] &=&[\wbf_{k,1}^{(M)}e^{j\varphi_{k,1}}, \
\wbf_{k,2}^{(M)}e^{j\varphi_{k,2}}, \ \cdots,\
\wbf_{k,d}^{(M)}e^{j\varphi_{k,d}}],\\
&=&[\underline{\Vbf}_k\phibf_k^{*(1)}e^{j\varphi_{k,1}},\
\underline{\Vbf}_k\phibf_k^{*(2)}e^{j\varphi_{k,2}},\ \cdots,\
\underline{\Vbf}_k\phibf_k^{*(d)}e^{j\varphi_{k,d}}],\\
&=&\underline{\Vbf}_k
\Phibf_k^* \mbox{diag}(e^{j\varphi_{k,1}},e^{j\varphi_{k,2}}, \cdots, e^{j\varphi_{k,d}}),\\
&=& [\vbf_k^{*(1)}e^{j\varphi_{k,1}},\
\vbf_k^{*(2)}e^{j\varphi_{k,2}},\
\cdots, ~\vbf_k^{*(d)}e^{j\varphi_{k,d}}]
\end{eqnarray}
for all $k$.  Now, recall from (\ref{eq:theorem4scalarwbf}) that
$r^\prime_{k,m}e^{j\varphi_{k,m}}=\wbf_{k,m}^{(M)H}\overleftarrow{\Hbf}_{kk}\ubf_{k}^{(m)}[n]$.
 Substituting
$\wbf_{k,m}^{(M)}=\vbf_k^{*(m)}$ and
$\ubf_{k}^{(m)}[n]=\ubf_k^{*(m)}e^{j\omega_{k,m}}$ into
(\ref{eq:theorem4scalarwbf}) yields
\begin{equation}
r^\prime_{k,m}e^{j\varphi_{k,m}}
=\wbf_{k,m}^{(M)H}\overleftarrow{\Hbf}_{kk}\ubf_{k}^{(m)}[n]
=\vbf_{k}^{*(m)H}\overleftarrow{\Hbf}_{kk}\ubf_{k}^{*(m)}e^{j\omega_{k,m}}.
\label{eq:thm5_phase5}
\end{equation}
Also, substituting $\qbf_{k,m}^{(M)}$ in (\ref{eq:thm4_phase1})
with
  (\ref{eq:theorem5qkiopt}) yields
\begin{equation}\label{eq:theorem5constantdeal}
r_{k,m}e^{j\omega_{k,m}}=\qbf_{k,m}^{(M)H}\Hbf_{kk}\vbf_k^{*(m)}=\ubf_k^{*(m)H}\Hbf_{kk}\vbf_k^{*(m)}.
\end{equation}
From  (\ref{eq:thm5_phase5}) and (\ref{eq:theorem5constantdeal})
we have
\begin{equation}
r^\prime_{k,m}e^{j\varphi_{k,m}}
=\vbf_{k}^{*(m)H}\overleftarrow{\Hbf}_{kk}\ubf_{k}^{*(m)}e^{j\omega_{k,m}}
=r_{k,m}e^{-j\omega_{k,m}}e^{j\omega_{k,m}}=r_{k,m},
\end{equation}
and thus $\varphi_{k,m}=0$ since both $r_{k,m}$ and
$r^\prime_{k,m}$ are real. This holds for all streams and users,
i.e., $\varphi_{k,m}=0$ for all $k$ and $m$. Finally, from
(\ref{eq:theorem5conclusionwithrotation}) and $\varphi_{k,m}=0$,
we have
\begin{equation}
\Vbf_{k}[n+1] =\Vbf_k^{*}=\Vbf_{k}[n].
\end{equation}
Thus, the two-layer linear beamforming solution
$\{\underbar{\Vbf}_k \Phibf_k^*, \underbar{\Ubf}_k\Thetabf_k^*\}$
is a fixed point of the max-SINR algorithm at sufficiently high
SNR.
\end{proof}

\vspace{0.5em} \noindent  Note that if the outer precoder and
decoder $\Phibf$ and $\Thetabf$ are not optimized to the single
user equivalent channel $\bar{\Hbf}_k$ resulting from interference
alignment by the inner precoder and decoder $\{\underbar{\Vbf}_k,
\underbar{\Ubf}_k\}$, then (\ref{eq:theorem5fixedPxbf}) and
(\ref{eq:theorem5qkiopt}) are not valid and thus the beamforming
matrices change with the iteration; {\em the interference-aligning
solution with  suboptimal outer coders (including the zero-forcing
$\Thetabf_k = (\bar{\Hbf}_k \Phibf)^\dagger$) within the given
space is not a fixed point of the max-SINR algorithm.} (Numerical
result confirming this will be shown in Section
\ref{sec:numerical}.) Now define the optimal interference-aligning
subspaces as the column spaces of matrices
\begin{equation}
\{\underbar{\Vbf}_k^*, \underbar{\Ubf}_k^* \}
 =\mathop{\arg\max}_{\underbar{\Vbf}_k, \underbar{\Ubf}_k} R(
\underbar{\Vbf}_k \Phibf_k^* (\{\underbar{\Vbf}_k,
\underbar{\Ubf}_k\}), \underbar{\Ubf}_k \Thetabf_k^*
(\{\underbar{\Vbf}_k, \underbar{\Ubf}_k\})),
\end{equation}
which achieves $R_{IA}^*$. Then, since Theorem
\ref{theo:TwoLayerFixedPointOfmaxSINR} is valid for any
interference-aligning matrices, we have the following corollary to
Theorem \ref{theo:TwoLayerFixedPointOfmaxSINR}.

\vspace{0.5em}
\begin{corollary} \label{corollary:optimalIAinnerouter}
$\{\Vbf_k^{**}=\underbar{\Vbf}_k^*
\Phibf^*(\underbar{\Vbf}_k^*,\underbar{\Ubf}_k^*),
\Ubf_k^{**}=\underbar{\Ubf}_k^*
\Thetabf^*(\underbar{\Vbf}_k^*,\underbar{\Ubf}_k^*)\}$ is a fixed
point of the max-SINR algorithm  at sufficiently high SNR.
\end{corollary}
\vspace{0.5em}

\begin{lemma} \label{lemma:GOFPmaxSINR} The globally optimal  fixed point
of the max-SINR algorithm (in the sense that it has the maximum
sum rate among all its fixed points) satisfies the interference
alignment condition at sufficiently high SNR.
\end{lemma}

\vspace{0.5em}
\begin{proof}
Suppose that the globally optimal fixed point does not satisfy the
interference alignment condition.  Then, due to the interference
leakage the SINR of a certain stream does not increase unboundedly
as the signal power tends to infinity and the maximum number of
degrees of freedom is not achieved  with this assumed  "globally
optimal" fixed point. Therefore, this assumed globally optimal
fixed point has lower sum rate than the interference-aligning
fixed point $\{\Vbf_k^{**}, \Ubf_k^{**}\}$ in Corollary
\ref{corollary:optimalIAinnerouter} at sufficiently high SNR, and
it is not globally optimal among the fixed points of the max-SINR
algorithm. Thus, we have a contradiction, and hence the claim
follows.
\end{proof}

\vspace{0.5em}
\begin{theorem} \label{theo:globallyoptimalfixedmaxSINR} $\{\Vbf_k^{**}, \Ubf_k^{**}\}$ in Corollary
\ref{corollary:optimalIAinnerouter} is the globally optimal fixed
point of the  max-SINR algorithm at sufficiently high SNR.
\end{theorem}

\vspace{0.5em}
\begin{proof}
By Lemma \ref{lemma:GOFPmaxSINR} the globally optimal fixed point
of the max-SINR algorithm is interference-aligning at sufficiently
high SNR. Among all interference-aligning beamforming matrices,
$\{\Vbf_k^{**}, \Ubf_k^{**}\}$ has the maximum sum rate $R_{IA}^*$
and it is also a fixed point of the max-SINR algorithm by
Corollary \ref{corollary:optimalIAinnerouter}. Hence,
$\{\Vbf_k^{**}, \Ubf_k^{**}\}$ is the globally optimal fixed-point
of the max-SINR algorithm at sufficiently high SNR.
\end{proof}

\vspace{1em} \noindent Theorem
\ref{theo:globallyoptimalfixedmaxSINR} established an optimality
property for the max-SINR algorithm at high SNR. At high SNR one
can deduce that the optimal beamformer among all linear
beamformers  should satisfy the interference alignment condition
at high SNR by an argument similar to that used in Lemma
\ref{lemma:GOFPmaxSINR} and thus $R_{IA}^*$ in
(\ref{eq:maxRateByIA}) is the best sum rate achievable by linear
processing at sufficiently high SNR. Thus, {\em the max-SINR
algorithm is optimal at high SNR among all linear beamformers in
the sense that the set of its fixed points includes the globally
optimal linear beamforming solution $\{\Vbf_k^{**},
\Ubf_k^{**}\}$.} One algorithmic advantage of the max-SINR
algorithm is that it updates the beamforming matrices at least to
be outer-coder optimal. The benefit is evident when we consider
the sum-rate gradient algorithm based on the user-by-user approach
(\ref{eq:theVeryDataModel}, \ref{eq:sumrategradCk},
\ref{eq:sumrategradTotalRate}), which has a much larger set of
fixed points as described in the following theorem.

\vspace{1em}
\begin{theorem} \label{theo:SumRateGradident}
Any set $\{\check{\Vbf}_k, k=1,2,\cdots,K\}$ of
interference-aligning beamforming matrices
 is a  fixed point of the sum-rate gradient
algorithm in Section \ref{subsubsec:SumRateAlgorithms} at
sufficiently high SNR.
\end{theorem}

\vspace{0.5em}
\begin{proof}
This result can be proven by computing the gradient
(\ref{eq:gradient}) at $\{\check{\Vbf}_k\}$. The gradient
(\ref{eq:gradient}) at $\{\check{\Vbf}_k\}$ is given by
\begin{equation} \label{eq:proofSumRateGradGrad1}
\nabla_{\Vbf_k^o} C(\Vbf_1,\cdots,\Vbf_K)|_{\check{\Vbf}_k}
=\sum_{l=1}^{K}
\frac{P_t}{Kd}\Hbf_{lk}^H\Rbf_l^{-1}\Hbf_{lk}\check{\Vbf}_k
-\sum_{l\neq k}
\frac{P_t}{Kd}\Hbf_{lk}^H(\Ibf+\Zbf_l)^{-1}\Hbf_{lk}\check{\Vbf}_k.
\end{equation}
The SVD of the term  $(\Ibf+\Zbf_l)^{-1}$ on the RHS of the above
equation is obtained from (\ref{eq:SVD_Zk}) and is given by
\begin{eqnarray}
(\Ibf+\Zbf_l)^{-1} &=& \left[ \zbf_l^{(1)}\ \zbf_l^{(2)}\ \cdots
\zbf_l^{(M)} \right] \left[ \begin{array}{cc}
\Sigmabf_l+\Ibf_{d\times d}  & {\mathbf{0}}_{d\times d} \\
{\mathbf{0}}_{d\times d}       & \Ibf_{d\times d}
\end{array} \right]^{-1}
\left[ \begin{array}{c}
\zbf_l^{(1)H} \\
\zbf_l^{(2)H} \\
\vdots \\
\zbf_l^{(M)H}
\end{array} \right]  \\
&=& \left[\zbf_l^{(d+1)}\ \zbf_l^{(d+2)}\ \cdots \zbf_l^{(M)}
\right] \left[\zbf_l^{(d+1)}\ \zbf_l^{(d+2)}\ \cdots \zbf_l^{(M)}
\right]^H \label{eq:proofSumRateGradGrad11}
\end{eqnarray}
as $P_t \rightarrow \infty$ since the nonzero singular values of
$\Zbf_l$ increases without bound as $P_t \rightarrow \infty$. From
Lemma \ref{lemma:UbfkTheSame}, we have
${\mathcal{C}}([\zbf_l^{(d+1)},\cdots,\zbf_l^{(M)}])=\Cc(\Zbf_l)^\perp
={\mathcal{C}}(\check{\Ubf}_l)$. Due to the interference alignment
by $\{\check{\Vbf}_k, \check{\Ubf}_k\}$, we have
\begin{equation} \label{eq:proofSumRateGradGrad2}
(\Ibf+\Zbf_l)^{-1}\Hbf_{lk}\check{\Vbf}_k={\mathbf{0}},
\end{equation}
 for $l\neq k$, and thus the second term on the RHS
of (\ref{eq:proofSumRateGradGrad1}) vanishes at sufficiently high
SNR.  Now consider the first term on the RHS of
(\ref{eq:proofSumRateGradGrad1}). From the matrix inversion lemma,
we have
\begin{eqnarray} \label{eq:thm4_relation}
\Rbf_l^{-1} &=&
\left(\Ibf+\Zbf_l+\frac{P_t}{Kd}\Hbf_{ll}\check{\Vbf}_l
\check{\Vbf}_l^{H}\Hbf_{ll}^H\right)^{-1} =
\left(\Mbf_l+\frac{P_t}{Kd}\Hbf_{ll}\check{\Vbf}_l
\check{\Vbf}_l^{H}\Hbf_{ll}^H\right)^{-1}\nonumber \\
&=& \Mbf_l^{-1} -\frac{P_t}{Kd}\Mbf_l^{-1}\Hbf_{ll}\check{\Vbf}_l
\left(\Ibf+\frac{P_t}{Kd}\check{\Vbf}_l^{H}\Hbf_{ii}^H
\Mbf_l^{-1}\Hbf_{ll}\check{\Vbf}_l
\right)^{-1}\check{\Vbf}_l^{H}\Hbf_{ll}^H\Mbf_l^{-1},
\end{eqnarray}
where $\Mbf_l = \Ibf + \Zbf_l$. Substituting this into the second
term yields
\begin{eqnarray}
&&\sum_{l=1}^{K}
\frac{P_t}{Kd}\Hbf_{lk}^H\Rbf_l^{-1}\Hbf_{lk}\check{\Vbf}_k  \\
&=& \sum_{l=1}^{K} \frac{P_t}{Kd}\Hbf_{lk}^H
\left(\Mbf_l^{-1}-\frac{P_t}{Kd}\Mbf_l^{-1}\Hbf_{ll}\check{\Vbf}_l
\left(\Ibf+\frac{P_t}{Kd}\check{\Vbf}_l^H\Hbf_{ll}^H\Mbf_l^{-1}\Hbf_{ll}
\check{\Vbf}_l\right)^{-1}
\check{\Vbf}_l^H\Hbf_{ll}^H\Mbf_l^{-1}\right)\Hbf_{lk}\check{\Vbf}_k.
\nonumber
\end{eqnarray}
Since $\Mbf_l^{-1}\Hbf_{lk}\check{\Vbf}_k={\mathbf{0}}$ for all
$l\neq k$ by (\ref{eq:proofSumRateGradGrad2}), we have
\begin{eqnarray}
&& \nabla_{\Vbf_k^o} C(\Vbf_1,\cdots,\Vbf_K)|_{\check{\Vbf}_k} =
\frac{P_t}{Kd}\Hbf_{kk}^H \Mbf_k^{-1} \Hbf_{kk}\check{\Vbf}_k \nonumber \\
& &-\left(\frac{P_t}{Kd}\right)^2 \Hbf_{kk}^H
\Mbf_k^{-1}\Hbf_{kk}\check{\Vbf}_k
\left(\Ibf+\frac{P_t}{Kd}\check{\Vbf}_k^{H}\Hbf_{kk}^H\Mbf_k^{-1}\Hbf_{kk}\check{\Vbf}_k\right)^{-1}\check{\Vbf}_k^{H}\Hbf_{kk}^H\Mbf_k^{-1}\Hbf_{kk}\check{\Vbf}_k
+o(1)\nonumber \\
&\stackrel{(a)}{=}& \frac{P_t}{Kd}\Hbf_{kk}^H\Mbf_k^{-1}\Hbf_{kk}\check{\Vbf}_k \nonumber \\
& &-\left(\frac{P_t}{Kd}\right)^2\Hbf_{kk}^H \Mbf_k^{-1}\Hbf_{kk}\check{\Vbf}_k \left(\frac{P_t}{Kd}\check{\Vbf}_k^{H}\Hbf_{kk}^H\Mbf_k^{-1}\Hbf_{kk}\check{\Vbf}_k \right)^{-1}\check{\Vbf}_k^{H}\Hbf_{kk}^H\Mbf_k^{-1}\Hbf_{kk}\check{\Vbf}_k +o(1)\nonumber \\
&=& \frac{P_t}{Kd}\Hbf_{kk}^H\Mbf_k^{-1}\Hbf_{kk}\check{\Vbf}_k
-\frac{P_t}{Kd}\Hbf_{kk}^H \Mbf_k^{-1}\Hbf_{kk}\check{\Vbf}_k + o(1)\nonumber \\
&=& {\mathbf{0}} ~~\mbox{as}~~ P_t \rightarrow \infty,
\end{eqnarray}
where (a) holds for sufficiently high SNR since $\Mbf_l^{-1}$ in
(\ref{eq:proofSumRateGradGrad11}) does not depend on $P_t$. Hence,
the gradient is zero at $\{\check{\Vbf}_k\}$ and
$\{\check{\Vbf}_k\}$ is a fixed point of the sum-rate gradient
algorithm at sufficiently high SNR.
\end{proof}

\vspace{1em} \noindent Now the difference between the
stream-by-stream approach and the user-by-user approach is clear
under equal power allocation to all streams. One could have
conjectured that aggregating all the streams of a user together
and formulating the sum rate problem correspondingly  to construct
an algorithm might yield better performance. However, there is an
algorithmic disadvantage of such a formulation at least at high
SNR. For the user-by-user approach (\ref{eq:theVeryDataModel},
\ref{eq:sumrategradCk}, \ref{eq:sumrategradTotalRate}) to
algorithm construction, there is no resolving power of the
algorithm to distinguish each stream of a user and thus the
algorithm yields only a DoF-optimal
 point at high SNR, i.e., it only yields a set of interference-aligning
 matrices (or user-by-user interference alignment), as shown in Theorem \ref{theo:SumRateGradident}. On the
 other hand, the max-SINR algorithm based on the stream-by-stream
 approach has the resolving capability to optimize each stream further
 and to yield at least a point with an optimal outer coder in addition to user-by-user interference alignment. Of course, the
 sum-rate gradient algorithm also contains $\{\Vbf_k^{**},
 \Ubf_k^{**}\}$ in its fixed-point set.

Finally, we examine the convergence behavior of the max-SINR
algorithm.  The conventional convergence analysis of the algorithm
is not straightforward due to its nonconvex nature and the
normalization step in each iteration to make  each beam vector
have norm one. Such a normalization is a projection to the surface
of a unit sphere in a high dimensional space and it is not a
non-expansive\footnote{A projection $\Pibf$ is called
non-expansive if $||\Pibf \xbf -\Pibf \xbf|| \le ||\xbf-\ybf||$
for all $\xbf$ and $\ybf$, which is a useful property for proving
convergence.} projection, unlike the non-expansive projection to a
unit sphere including the inside \cite{Stark&Yang:book}, which
makes the application of general convergence analysis tools
difficult. To circumvent this difficulty, we consider the local
convergence since we have already shown the existence of a fixed
point. Based on the perturbation approach, we provide the local
convergence behavior of the max-SINR algorithm at high SNR, given
in the following theorem.

\vspace{1em}
\begin{theorem} \label{theo:LocalConvergence}
The max-SINR algorithm converges to a fixed point exponentially
when it is initialized within a neighborhood around the fixed
point at sufficiently high SNR.
\end{theorem}

\vspace{0.5em}
\begin{proof}
Any point $\tilde{\Vbf}_k$ within an $\epsilon$-neighborhood of a
fixed point $\Vbf^*$ is represented as
\begin{equation}
\tilde{\Vbf}_k=(\Vbf_k^* + \epsilon\Pbf_k ) \left[
\begin{array}{ccccc}
\alpha_k^{(1)} & & & \\
& \alpha_k^{(2)} & & \\
& & \ddots & \\
& & & \alpha_k^{(d)}
\end{array} \right]
=(\Vbf_k^*+\epsilon\Pbf_k)\Abf_k,
\end{equation}
where the matrix $\Pbf_k$ consists of arbitrary unit-norm vectors.
Here $\{\alpha_k^{(m)}\}$ are the normalization factors so that
each column of the initial point $\tilde{\Vbf}_k$ has unit norm.
Since both $\Vbf_k^*$ and $\Pbf_k$ have unit-norm column vectors,
we have $1 -\epsilon \le \alpha_k^{(m)} \le 1 + \epsilon$. Thus,
we have $\Abf_k=\Ibf-\epsilon\Dbf_k$, where
$\Dbf_k=\mbox{diag}(\eta_k^{(1)},\eta_k^{(2)},\cdots,\eta_k^{(d)})$
and $-1\le \eta_k^{(m)} \le 1$. The interference-plus-noise
covariance matrix with the initialization $\tilde{\Vbf}_k$ is
given by
\begin{eqnarray}
\tilde\Rbf_k^{(m)} &=& \sum_{l\neq
k}\frac{P_t}{Kd}\Hbf_{kl}\tilde\Vbf_l
   \tilde\Vbf_l^H\Hbf_{kl}^H
 +\sum_{j\neq m}\frac{P_t}{Kd}\Hbf_{kk}\tilde\vbf_k^{(j)}
 \tilde\vbf_k^{(j)H}\Hbf_{kk}^H + \Ibf, \nonumber\\
    &=& \sum_{l\neq
k}\frac{P_t}{Kd}\Hbf_{kl}(\Vbf_l^*+\epsilon\Pbf_l)
    (\Ibf-\epsilon\Dbf_l)^2 (\Vbf_l^*+\epsilon\Pbf_l)^H\Hbf_{kl}^H
    \nonumber\\
     & & ~~~~~+\sum_{j\neq m}\frac{P_t}{Kd}
(1-\epsilon\eta_k^{j})^2 \Hbf_{kk}
    (\vbf_k^{*(j)}+\epsilon{\mathbf{p}}_k^{(j)})
    (\vbf_k^{*(j)}+\epsilon{\mathbf{p}}_k^{(j)})^H\Hbf_{kk}^H +
    \Ibf.
     \label{eq:tildeRbfkm}
\end{eqnarray}
After some manipulation, (\ref{eq:tildeRbfkm}) is given by
\begin{equation}
\tilde\Rbf_k^{(m)} = \Rbf_k^{(m)}+\epsilon
\Delta\Rbf_k^{(m)}+o(\epsilon),
\end{equation}
where $\Rbf_k^{(m)}$ is the interference-plus-noise covariance
matrix based on the fixed point $\Vbf_k^*$ and
\begin{eqnarray}
\Delta\Rbf_k^{(m)} &=& \sum_{l\neq k}\frac{P_t}{Kd} \Hbf_{kl}
    (\Vbf_l^*\Pbf_l^H+\Pbf_l\Vbf_l^{*H})\Hbf_{kl}^H
    +\sum_{j\neq m}\frac{P_t}{Kd}\Hbf_{kk}
    (\vbf_k^{*(j)}{\mathbf{p}}_k^{(j)H}
    +{\mathbf{p}}_k^{(j)}\vbf_k^{*(j)H})\Hbf_{kk}^H  \nonumber \\
& &\qquad\quad  -2  \sum_{l\neq k}
     \frac{P_t}{Kd}\Hbf_{kl}\Vbf_l^*\Dbf_l\Vbf_l^{*H}\Hbf_{kl}^H
    -2\sum_{j\neq m}\eta_k^{(j)}\frac{P_t}{Kd}\Hbf_{kk}\vbf_k^{*(j)}
     \vbf_k^{*(j)H}\Hbf_{kk}^H. \label{eq:theoConvDeltaRbfkm}
\end{eqnarray}
Note that $\Delta\Rbf_k^{(m)}$ is a full rank matrix unless
$\Pbf_l =  \Vbf_l^*$ since the third and fourth terms of the RHS
of (\ref{eq:theoConvDeltaRbfkm}) already yield rank of $M-1$ and
$\vbf_k^{*(j)}{\mathbf{p}}_k^{(j)H}$ adds one to the total rank.
(In case of $\Pbf_l =  \Vbf_l^*$, simply $\Delta\Rbf_k^{(m)}$ is a
scaled version of $\Rbf_k^{(m)} -\Ibf$.)  Applying the matrix
inversion lemma to $\tilde{\Rbf}_k^{(m)}$ successively, we have
\begin{eqnarray}
(\tilde\Rbf_k^{(m)})^{-1}
&=& (\Rbf_k^{(m)}+\epsilon\Delta\Rbf_k^{(l)})^{-1}+o(\epsilon), \nonumber \\
&=& (\Rbf_k^{(m)})^{-1}-\epsilon(\Rbf_k^{(m)})^{-1}
\left\{(\Delta\Rbf_k^{(m)})^{-1}+\epsilon(\Rbf_k^{(m)})^{-1}\right\}^{-1}(\Rbf_k^{(m)})^{-1}+o(\epsilon), \nonumber \\
&=&
 (\Rbf_k^{(m)})^{-1}-\epsilon(\Rbf_k^{(m)})^{-1} \Xbf_k
 (\Rbf_k^{(m)})^{-1}+o(\epsilon),
 \label{eq:theoConvTildeRbfkmInv}
\end{eqnarray}
where
$\Xbf_k=\left\{(\Delta\Rbf_k^{(m)})^{-1}+\epsilon(\Rbf_k^{(m)})^{-1}\right\}^{-1}$.
  Thus, the
unnormalized receive beamforming vector by the VU update is given
by
\begin{equation} \label{eq:theoConvTildeUbfKm}
\tilde\ubf_k^{(m)}
=(\tilde\Rbf_k^{(m)})^{-1}\Hbf_{kk}\tilde{\vbf}_k^{(m)} =(\tilde\Rbf_k^{(m)})^{-1}\Hbf_{kk}(\vbf_k^{*(m)}+\epsilon{\mathbf{p}}_k^{(m)})(1-\epsilon\eta_k^{(m)}). \\
\end{equation}
Applying (\ref{eq:theoConvTildeRbfkmInv}) to
(\ref{eq:theoConvTildeUbfKm}), we have after some manipulation
{\small
\begin{eqnarray}
\tilde\ubf_k^{(m)}&=& (\Rbf_k^{(m)})^{-1}\Hbf_{kk}\vbf_k^{*(m)}
\label{eq:theolocalconvtildeubfkm}
\\
&&
-\epsilon\left\{\eta_k^{(m)}(\Rbf_k^{(m)})^{-1}\Hbf_{kk}\vbf_k^{*(m)}
+(\Rbf_k^{(m)})^{-1}\Xbf_k(\Rbf_k^{(m)})^{-1}\Hbf_{kk}\vbf_k^{*(m)}
-(\Rbf_k^{(m)})^{-1}\Hbf_{kk}{\mathbf{p}}_k^{(m)}\right\}
+o(\epsilon), \nonumber
\end{eqnarray}
} where $o(\epsilon)$ term starts with $\epsilon^2$ order. From
(\ref{eq:theorem4_svd_q}) we have
\begin{equation}
(\Rbf_k^{(m)})^{-1} =\qbf_{k,m}^{(M)}\qbf_{k,m}^{(M)H} +
\sum_{i=1}^{M-1}
\frac{1}{1+\lambda_{k,m}^{(i)}}\qbf_{k,m}^{(i)}\qbf_{k,m}^{(i)H}.
\end{equation}
As the SNR increases, $\lambda_{k,m}^{(i)} \rightarrow \infty$ and
we have
\begin{equation}  \label{eq:theolocalconvTraceNorm}
(\Rbf_k^{(m)})^{-1} = \qbf_{k,m}^{(M)}\qbf_{k,m}^{(M)H} + \delta
\Qbf_{k,m},
\end{equation}
for arbitrary $\delta$, where $\Qbf_{k,m}$ has the trace norm less
than one. Substituting (\ref{eq:theolocalconvTraceNorm}) into
(\ref{eq:theolocalconvtildeubfkm}) yields
\begin{equation}
\tilde\ubf_k^{(m)} = \qbf_{k,m}^{(M)}c_1 + \delta \Qbf_{k,m}\cbf_1
-\epsilon\left\{\qbf_{k,m}^{(M)} c_2 + \delta \Qbf_{k,m} \cbf_{2}
\right\} + \epsilon^2 + o(\epsilon^2),
\end{equation}
where $c_1$ and $c_2$ are finite constants and $\cbf_1$ and
$\cbf_2$ are vectors with finite norm. Since $\delta$ is arbitrary
at high SNR, the terms with $\delta$ are negligible. The
$\epsilon$-linear perturbation output  is aligned with
$\qbf_{k,m}^{(M)}$ which is $\ubf_k^{*(m)}$ at high SNR. (See
Theorem \ref{theo:TwoLayerFixedPointOfmaxSINR}.)  Thus, the linear
perturbation term in $\epsilon$ disappears by being aligned to the
fixed point vector and after this the high order terms starting
from the second order in $\epsilon$ remain. Since the overall
normalization is on $\tilde{\ubf}_k^{(m)}$, it does not change the
relative size of each term.  Similarly, the elimination of the
dominant perturbation term holds to the UV step since the UV step
is the same as the VU step only with the role exchange between
$\Ubf_k$ and $\Vbf_k$.  Thus, the algorithm converges to the fixed
point.
\end{proof}

\vspace{0.5em} \noindent Now the behavior of the max-SINR
algorithm  is clear at least at high SNR. When it is near a fixed
point, the algorithm converges exponentially; at every step it
converges to the fixed point by eliminating the perturbation by
aligning to the fixed point by factor $\epsilon$.

\section{Numerical Results}
\label{sec:numerical}

In this section, we provide some numerical results to validate our
analysis in the previous sections. We considered the sum rate
performance of several beamformer design methods: two layer
optimal inner and outer precoder/decoder design based on
interference alignment and channel diagonalization, two-layer
suboptimal design with interference-aligning inner beamforming and
zero-forcing outer filter, i.e., $\Thetabf_k =
(\bar{\Hbf}_k\Phibf_k)^\dagger$ and $\Phibf_k = \Ibf$ in
(\ref{eq:optimalbeam11}), and the max-SINR algorithm with
orthogonalization. (The orthogonalization step will be explained
shortly.) First, we randomly generated a set of MIMO channel
matrices $\{\Hbf_{kl}\}$. For this given channel realization, each
algorithm was run 500 times with different random initialization
(hoping to converge to different fixed points). Here, we used the
iterative interference alignment (IIA) algorithm in
\cite{Gomadam&Jafar:08ArxiV} to obtain the interference-aligning
subspaces.

\begin{figure}[htbp]
\centerline{ \SetLabels
\L(0.25*-0.1) (a) \\
\L(0.76*-0.1) (b) \\
\endSetLabels
\leavevmode
\strut\AffixLabels{
\scalefig{0.5}\epsfbox{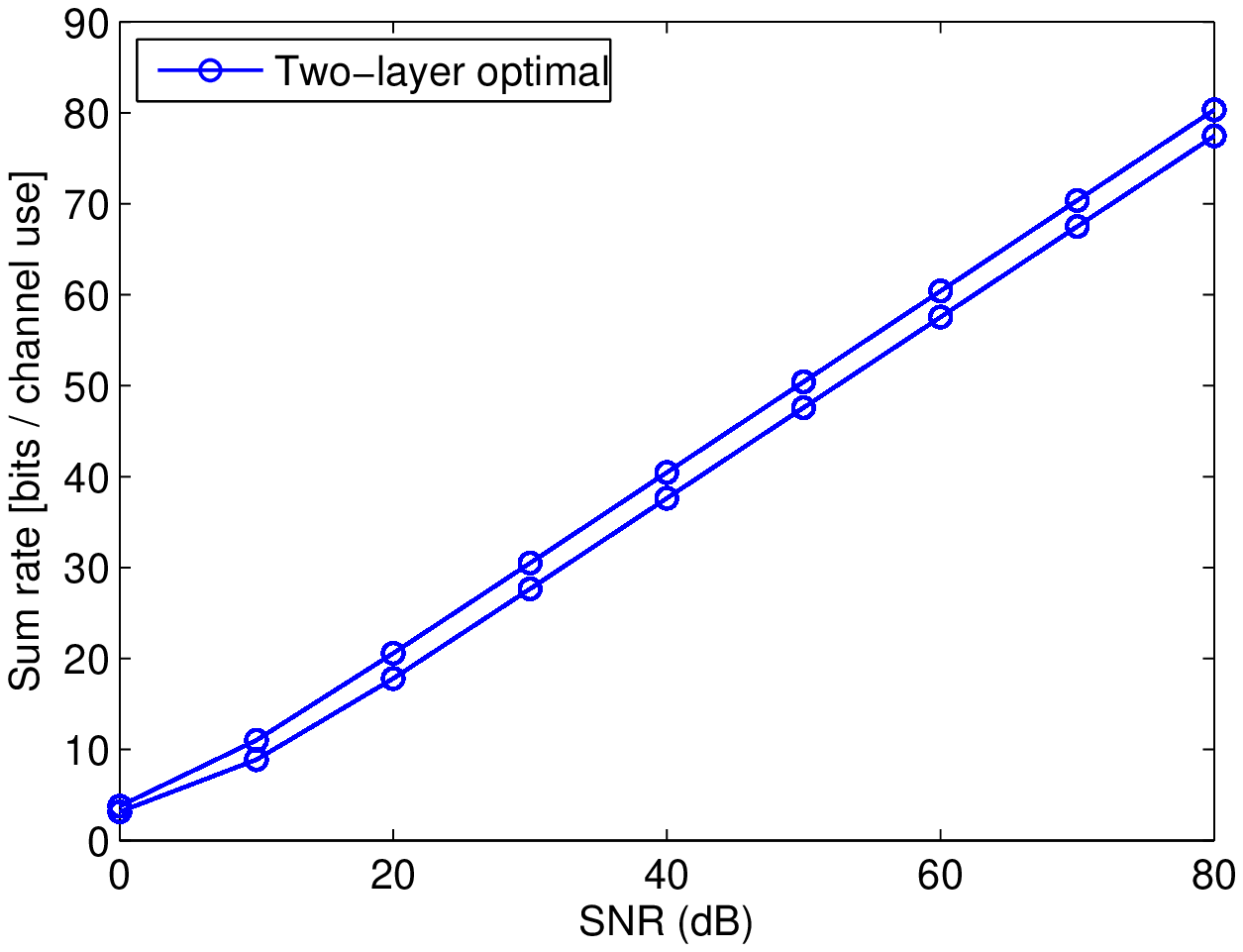}
\scalefig{0.5}\epsfbox{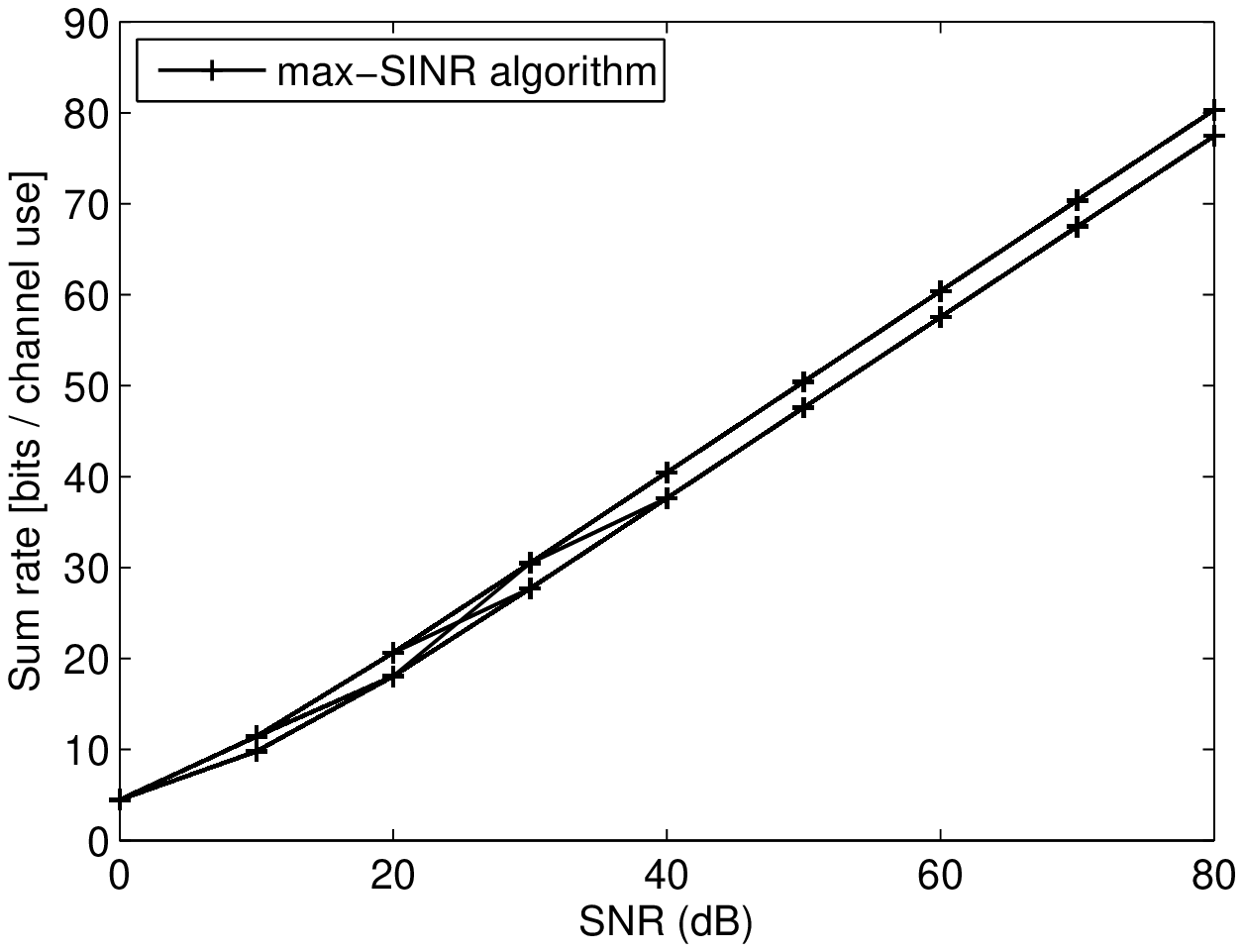} }
} \vspace{1cm} \centerline{ \SetLabels
\L(0.25*-0.1) (c) \\
\L(0.76*-0.1) (d) \\
\endSetLabels
\leavevmode
\strut\AffixLabels{
\scalefig{0.5}\epsfbox{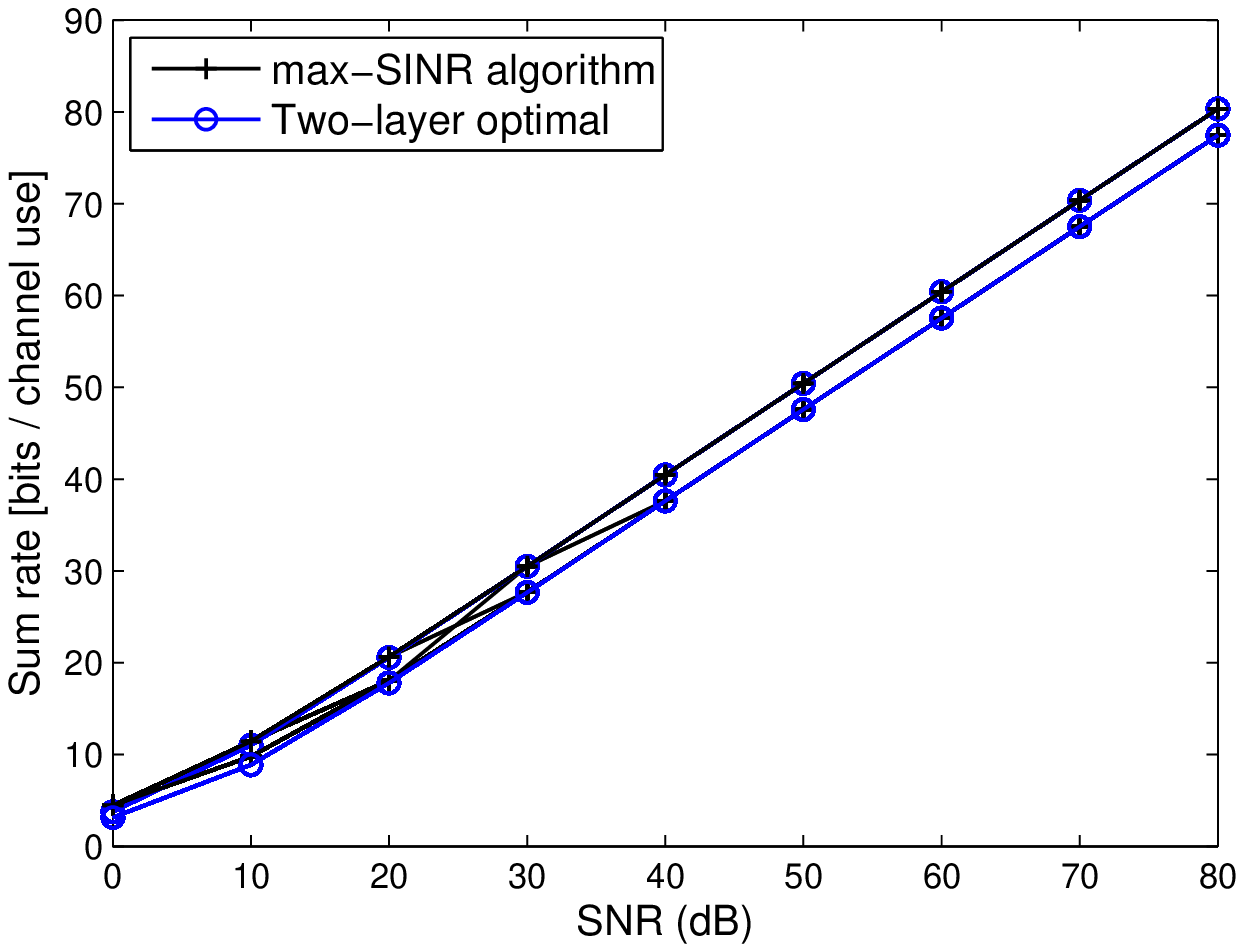}
\scalefig{0.5}\epsfbox{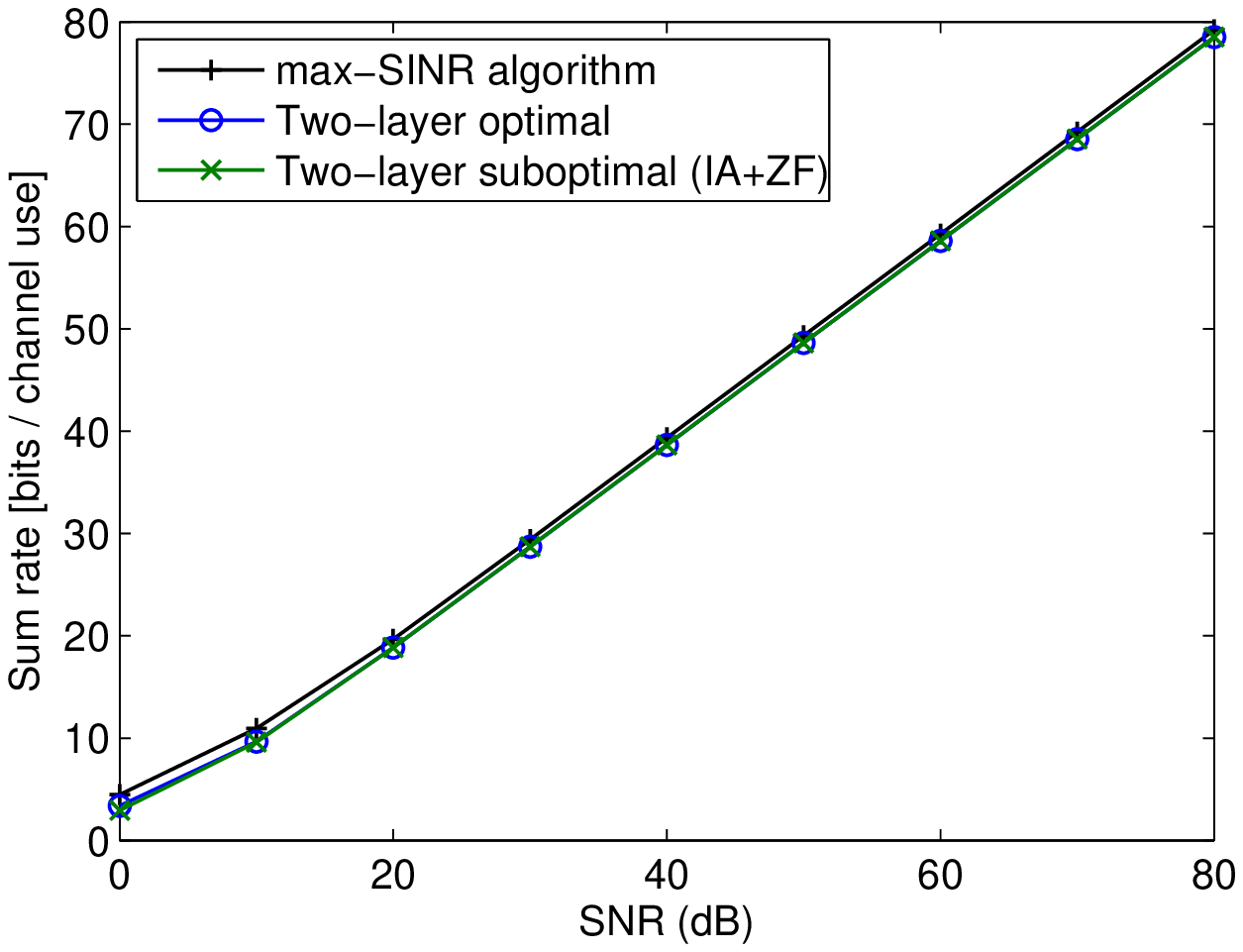} }
} \vspace{1cm} \caption{Sum rate as a function of  SNR with $500$
random initializations: (a) interference alignment with optimal
power allocation, (b) max-SINR algorithm, (c) (a) and (b) in a
single figure, (d) sum rate averaged over initializations}
\label{fig:Single_channel_N2M2d1}
\end{figure}

\begin{center}
\begin{table}
{\scriptsize
\begin{tabular}{|c|c|c|c|c|c|c|c|c|c|c|}
\hline
Algorithm &  & 0dB & 10dB & 20dB & 30dB & 40dB & 50dB & 60dB & 70dB & 80dB \\
\hline \hline
\multirow{5}{*}{max-SINR algorithm} & \multirow{2}{*}{F1} & 4.46  & 11.43 & 20.64 & 30.51 & 40.47 & 50.43  & 60.40 & 70.37 & 80.33 \\
 &  & (100) & (71) & (62)  & (61) & (61) & (61) & (61) & (61) & (61)\\ \cline{2-11}

& \multirow{2}{*}{F2}  & -  & 9.79 & 18.05 & 27.68 & 37.61 & 47.57 & 57.54  & 67.51 & 77.47\\
 &  & (0) & (29) & (38)  & (39) & (39) & (39) & (39) & (39) & (39)\\ \cline{2-11}
 &  Average rate  & 4.46 &10.95 & 19.65 & 29.42 & 39.35 & 49.32 & 59.28 & 69.25 & 79.22 \\
\hline\hline

\multirow{5}{*}{IA with optimal power allocation} & \multirow{2}{*}{F1} & 3.78 & 11.02 & 20.58 &  30.51 & 40.47 & 50.43 & 60.40 & 70.37 & 80.33 \\
 &  & (37) & (37) & (37) & (37)  & (37) & (37) & (37) & (37) & (37) \\ \cline{2-11}

& \multirow{2}{*}{F2}  & 3.14 & 8.86 & 17.81 & 27.66  & 37.61  & 47.57 & 57.54  & 67.51 & 77.47 \\
 &  & (63) & (63) & (63) & (63) & (63) & (63) & (63) & (63) & (63) \\ \cline{2-11}
 &  Average rate  & 3.38 & 9.66 & 18.83 & 28.71 & 38.67 & 48.63 & 58.60 & 68.56 & 78.53 \\
\hline\hline

\multirow{5}{*}{IA with equal power allocation} & \multirow{2}{*}{F1} & 3.53 & 11.01 & 20.58  &  30.51 & 40.47 & 50.43 & 60.40 & 70.37 & 80.33  \\
 &  & (37) & (37) & (37) & (37) & (37) & (37) & (37) & (37) & (37)\\ \cline{2-11}
 & \multirow{2}{*}{F2}  & 2.59 & 8.75 & 17.80 & 27.66 & 37.61 & 47.57 & 57.54 & 67.51 & 77.47 \\
 &  & (63) & (63) & (63) & (63) & (63) & (63) & (63) & (63)  & (63) \\ \cline{2-11}
 &  Average rate  & 2.94 & 9.59 & 18.83 & 28.71 & 38.67 & 48.63 & 58.60 & 68.56 & 78.53 \\
\hline
\end{tabular}
\caption{Sum rate and percentage of fixed points  as  functions of
SNR when $M=2$ and $d=1$: the numbers without and with parenthesis
in each box correspond to sum rate and percentage of that fixed
point over 500 random initializations, respectively.} }
\label{table:Single_channel_N2M2d1}
\end{table}
\end{center}

 Fig. \ref{fig:Single_channel_N2M2d1} shows the sum rate performance
in case of $K=3$ and $M=2d=2$. In this case, there is no outer
coder issue since $d=1$, but we can optimize power allocation
under interference alignment. Fig. \ref{fig:Single_channel_N2M2d1}
(a) shows the sum rate curve as a function of SNR for all
initializations for interference alignment with optimal power
allocation. (The 500 curves overlap on the two distinct curves.
Here, we generated one initialization randomly, and the IIA
algorithm was run for different SNR values with the same
initialization. This procedure was repeated over 500 times.) From
\cite{Cadambe&Jafar:08IT} we know the number of
interference-aligning subspaces is {\scriptsize $\left(
\begin{array}{c}  M \\ d \end{array} \right)$} for $M \times M$
MIMO channels when interference alignment is feasible. Indeed, we
see in the figure that there exist only two modes or fixed points
in $2 \times 2$ MIMO with $d=1$. Since the IIA algorithm does not
depend on the SNR, if one initialization ends with either of the
two fixed points, that initialization ends with the same fixed
point regardless of the value of SNR, i.e., we see two parallel
lines in the figure. On the other hand,  Fig.
\ref{fig:Single_channel_N2M2d1} (b) shows the sum rate curve as a
function of SNR for all initializations for the max-SINR
algorithm. (The procedure was the same as that of Fig.
\ref{fig:Single_channel_N2M2d1} (a).) Since the algorithm depends
on the SNR, now the surface of the sum rate functional changes
with SNR and the same initialization can lead to a different fixed
point as SNR changes. That is, there are the cross-over lines in
Fig. \ref{fig:Single_channel_N2M2d1} (b). It was also observed
that the cross over is unidirectional in most cases, i.e., we only
have either an $F1$ to $F2$ or an $F2$ to $F1$ transition once in
most cases as SNR increases. We seldom have  such a transition as
$F1\rightarrow F2 \rightarrow F1$, which implies that the sum-rate
surface is not too irregular.
 Fig. \ref{fig:Single_channel_N2M2d1} (c) is the figure combining Fig.
\ref{fig:Single_channel_N2M2d1} (a) and (b) together. It is seen
in the figure that for each mode (or fixed point) the max-SINR
algorithm and interference alignment coincide at high SNR, as
predicted by Theorem \ref{theo:TwoLayerFixedPointOfmaxSINR}. Table
\ref{table:Single_channel_N2M2d1} shows more detailed information
in this case. At low SNR, there is some gain to be had by optimal
power allocation for the same interference alignment, but the gain
is negligible and the two coincide as SNR increases since equal
power allocation is optimal at high SNR. As expected, the max-SINR
algorithm has further gain over interference alignment with
optimal power allocation at low SNR (e.g. 0 dB) for each mode.
 At high SNR, all three coincide for each mode as predicted.  One
interesting observation is that the interference alignment and the
max-SINR algorithm have almost the same sum rate in each mode from
10 dB SNR and the difference in the average sum rate in Fig.
\ref{fig:Single_channel_N2M2d1} (d) comes from the fact that the
max-SINR algorithm is more likely to end with the better fixed
point (i.e., the globally optimal one, F1, in this case) with
arbitrary initialization than is interference alignment across all
the SNR values.

\begin{center}
\begin{table}
{\scriptsize
\begin{tabular}{|c|c|c|c|c|c|c|c|c|c|c|}
\hline
Algorithm &  & 0dB & 10dB & 20dB & 30dB & 40dB & 50dB & 60dB & 70dB & 80dB \\
\hline \hline \multirow{13}{8em}{max-SINR with orthogonalization}
       & \multirow{2}{*}{F1} & 8.91 & 20.88 & 36.41 & 54.75 & 74.53 & 94.45 & 114.38 & 134.31 & 154.24 \\
          &   & (100) & (15.4) & (16.0) & (35.6) & (29.2) & (28.0) & (28.0) & (28.0) & (28.0) \\  \cline{2-11}

       & \multirow{2}{*}{F2} & - & 20.82 & 35.90 & 54.45 & 74.12 & 94.03 & 113.95 & 133.89 & 153.82 \\
          &   & (0) & (43.4) & (41.6) & (13.4) & (13.0) & (13.0) & (13.0) & (13.0) & (13.0) \\  \cline{2-11}

       & \multirow{2}{*}{F3} & - & 20.55 & 35.88 & 53.58 & 73.18 & 93.10 & 113.03 & 132.97 & 152.90 \\
           &   & (0) & (6.6) & (35.6) & (27.8) & (0.8) & (0.8) & (0.8) & (0.8) & (0.8) \\  \cline{2-11}

       & \multirow{2}{*}{F4} & - & 20.42 & 34.24 &  52.34 & 72.14 & 91.68 & 111.55 & 131.48 & 151.41 \\
           &   & (0) & (27.4) & (0.6)  & (0.8)    & (33.0) & (34.4) & (34.2) & (34.4) & (34.4) \\  \cline{2-11}

       & \multirow{2}{*}{F5} & - & 20.41 & 34.23 & 52.132 & 71.62 & 91.50 & 111.42 & 131.35 & 151.29 \\
          &   & (0) & (7.2) & (6.2) & (11.4) & (12.4) & (12.4) & (12.4) & (12.4) & (12.4) \\  \cline{2-11}

       & \multirow{2}{*}{F6} & - & - & - & 52.130 & 71.41 & 91.26 & 111.18 & 131.11 & 151.04 \\
          &   & (0) & (0) & (0) & (11.0) & (11.6) & (11.4) & (11.6) & (11.4) & (11.4) \\  \cline{2-11}

         &  Average rate  & 8.91 & 20.67 & 35.86 & 53.79 & 72.96 & 92.70 & 112.61 & 132.54 & 152.47 \\
\hline\hline

\multirow{13}{*}{Two-layer optimal}
       & \multirow{2}{*}{F1} & 7.21 & 19.18 & 35.08 & 54.63 & 74.52 & 94.45 & 114.38 & 134.31 & 154.24 \\
          &   & (33.2) & (33.2) & (33.2) & (13.8) & (13.8) & (13.8) & (13.8) & (13.8) & (13.8) \\  \cline{2-11}

       & \multirow{2}{*}{F2} & 7.16 & 18.14 & 35.08 & 54.19 & 74.09 & 94.02 & 113.95 & 133.89 & 153.82 \\
          &   & (12.0) & (13.8) & (13.8) & (9.8) & (9.8) & (9.8) & (9.8) & (9.8) & (9.8) \\  \cline{2-11}

       & \multirow{2}{*}{F3} & 6.93 & 17.95 & 34.56 & 53.27 & 73.17 & 93.10 & 113.03 & 132.97 & 152.90 \\
           &   & (26.0) & (12.0) & (9.8) & (5.2) & (5.2) & (5.2) & (5.2) & (5.2) & (5.2) \\  \cline{2-11}

       & \multirow{2}{*}{F4} & 6.82 & 17.76 & 33.63 & 52.27 & 71.74 & 91.62 & 111.55 & 131.48 & 151.41 \\
           &   & (13.8) & (26.0) & (5.2) & (33.2) & (33.2) & (33.2) & (33.2) & (33.2) & (33.2) \\  \cline{2-11}

       & \multirow{2}{*}{F5} & 6.58 & 17.18 & 32.67 & 51.73 & 71.57 & 91.49 & 111.42 & 131.35 & 151.29 \\
          &   & (9.8) & (9.8) & (12.0) & (12.0) & (12.0) & (12.0) & (12.0) & (12.0) & (12.0) \\  \cline{2-11}

       & \multirow{2}{*}{F6} & 5.50 & 16.15 & 32.52 & 51.50 & 71.33 & 91.25 & 111.18 & 131.11 & 151.04 \\
          &   & (5.2) & (5.2) & (26.0) & (26.0) & (26.0) & (26.0) & (26.0) & (26.0) & (26.0) \\  \cline{2-11}

         &  Average rate  & 6.93 & 18.17 & 34.00 & 52.57 & 72.30 & 92.21 & 112.14 & 132.07 & 152.00 \\
\hline \hline

\multirow{13}{*}{Two-layer suboptimal}
       & \multirow{2}{*}{F1} & 4.21 & 15.55 & 33.31 & 52.96 & 72.86 & 92.79 & 112.72 & 132.65 & 152.58 \\
          &   & (5.2) & (5.2) & (5.2) & (5.2) & (5.2) & (5.2) & (5.2) & (5.2) & (5.2) \\  \cline{2-11}

       & \multirow{2}{*}{F2} & 4.05 & 14.58 & 31.80 & 51.36 & 71.25 & 91.18 & 111.11 & 131.04 & 150.97 \\
          &   & (13.8) & (13.8) & (13.8) & (13.8) & (13.8) & (13.8) & (13.8) & (13.8) & (13.8) \\  \cline{2-11}

       & \multirow{2}{*}{F3} & 3.56 & 12.94 & 29.34 & 48.97 & 68.87 & 88.80 & 108.73 & 128.66 & 148.59 \\
           &   & (33.2) & (33.2) & (9.8) & (9.8) & (9.8) & (9.8) & (9.8) & (9.8) & (9.8) \\  \cline{2-11}

       & \multirow{2}{*}{F4} & 2.82 & 11.88 & 26.96 & 44.52 & 64.02 & 83.90 & 103.83 & 123.76 & 143.69 \\
           &   & (12.0) & (9.8) & (33.2) & (33.2) & (33.2) & (33.2) & (33.2) & (33.2) & (33.2) \\  \cline{2-11}

       & \multirow{2}{*}{F5} & 2.65 & 10.18 & 25.04 & 44.13 & 63.97 & 83.89 & 103.82 & 123.75 & 143.68 \\
          &   & (26.0) & (12.0) & (12.0) & (12.0) & (12.0) & (12.0) & (12.0) & (12.0) & (12.0) \\  \cline{2-11}

       & \multirow{2}{*}{F6} & 2.36 & 10.14 & 24.94 & 43.96 & 63.79 & 83.71 & 106.64 & 123.58 & 143.51 \\
          &   & (9.8) & (26.0) & (26.0) & (26.0) & (26.0) & (26.0) & (26.0) & (26.0) & (26.0) \\  \cline{2-11}

         &  Average rate  & 3.22 & 12.14 & 27.44 & 46.15 & 65.89 & 85.80 & 105.73 & 125.66 & 145.59 \\
\hline
\end{tabular}
\caption{Sum rate and percentage of fixed points as functions of
SNR when $M=4$ and $d=2$} } \label{table:Single_channel_N4M4d2}
\end{table}
\end{center}

\begin{figure}[ht]
\centering
\scalefig{0.8}\epsfbox{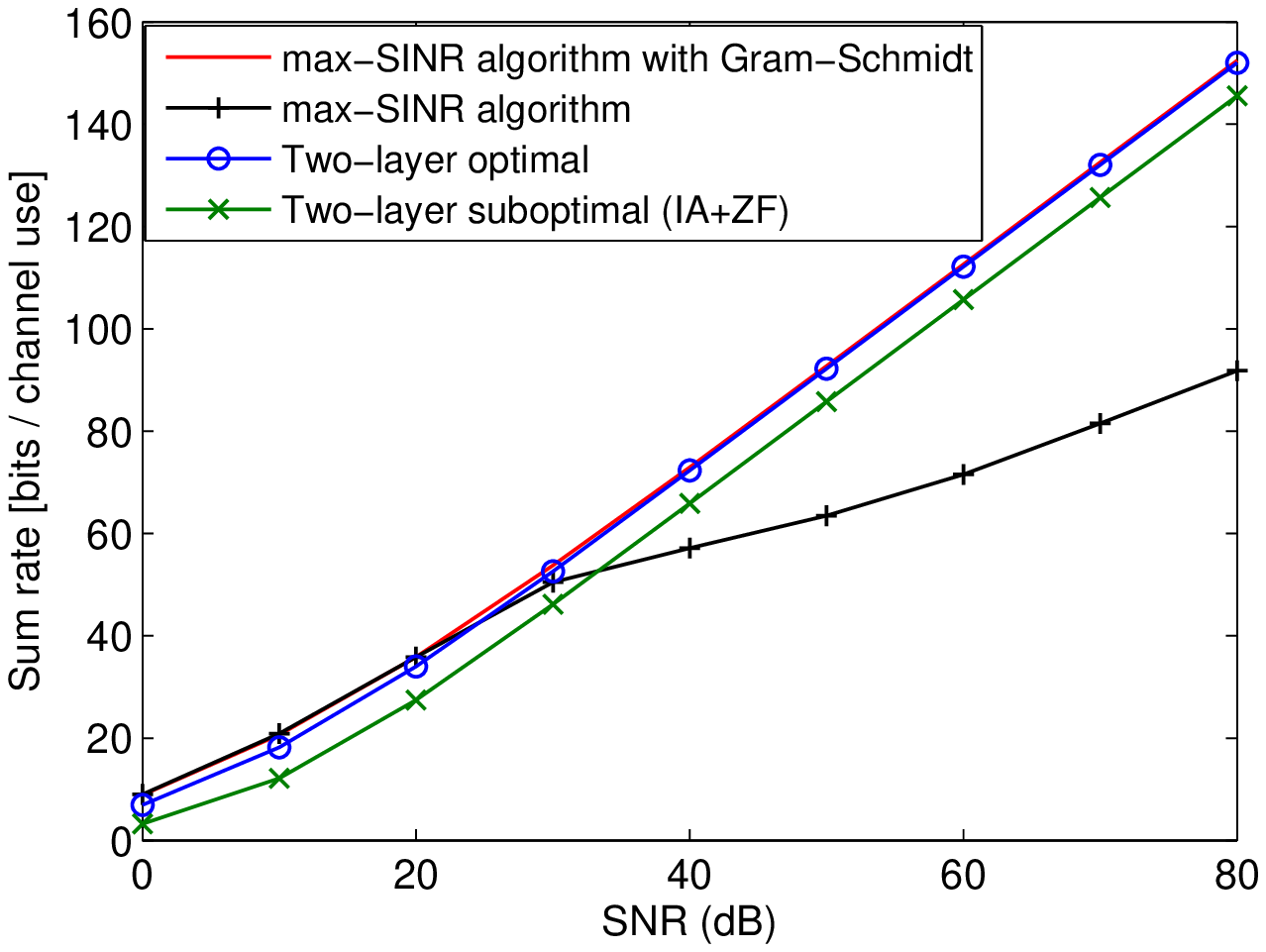}
\caption{Average sum rate when $K=3$, $M=4$ and $d=2$}
\label{fig:single_channel_N4M4d2}
\end{figure}

In case of $M=4$ and $d=2$, on the other hand, the situation is
more complicated.  Table \ref{table:Single_channel_N4M4d2} shows
the sum rate performance in this case. The same procedure was
performed as in the single stream case.  We observed {\scriptsize
$\left(
\begin{array}{c}  4 \\ 2 \end{array} \right)$}$=6$ fixed points\footnote{
In the two-layer suboptimal case, each point in the same
interference-aligning subspace may yield a different sum rate and
the number of observed sum rates might not have been six. However,
if we confine to beamforming matrices with columns' being an
orthonormal basis for the subspace, then the sum rate is the same
for the subspace regardless of the choice of the orthonormal
basis. That is, $C_k = \log | \Ibf + \Ubf^H \Hbf_{kk}\Vbf \Vbf^H
\Hbf_{kk}^H \Ubf| = \log | \Ibf + \tilde{\Ubf}^H
\Hbf_{kk}\tilde{\Vbf} \tilde{\Vbf}^H \Hbf_{kk}^H\tilde{\Ubf}|$ if
$\Ubf$ and $\tilde{\Ubf}$ are two different orthogonal bases for
the same subspace and so are $\Vbf$ and $\tilde{\Vbf}$. This is
the case with the IIA algorithm which returns beamforming matrices
with orthonormal  columns, and thus we observe six sum rate values
in this case also.} at each SNR, denoted as F1 to F6 in the table.
Now the role of the outer precoder and decoder is clearly seen.
Comparing the optimal two-layer and suboptimal two-layer designs
based on interference alignment, we see that there is noticeable
degradation in the suboptimal outer coder case with the same
interference-aligning subspace, i.e., the same mode. The gap does
not vanish as SNR increases. Also, the table shows that the
max-SINR algorithm indeed coincides with the two-layer optimal
beamforming of Section \ref{sec:optimaldesign} for each of the six
modes at high SNR, predicted by our Theorem
\ref{theo:TwoLayerFixedPointOfmaxSINR}; all the fixed points of
the max-SINR algorithm are two-layer optimal points at high SNR.
The F1's of the two algorithms are the same globally optimal
linear beamforming point, i.e., $\{\Vbf_k^{**}, \Ubf_k^{**}\}$ in
Theorem \ref{theo:globallyoptimalfixedmaxSINR}. Consistently with
the case of $d=1$, it is seen that the max-SINR algorithm is more
likely to converge to a better fixed point. (See the distribution
of fixed points.) Another interesting fact is that at 0 dB SNR the
max-SINR algorithm shows only one mode. This can easily be
explained. The term $\Rbf_{k}^{(m)}$ on the RHS of
(\ref{eq:individualRate}) converges to $\Ibf$ at very low SNR and
$C_k^{(m)}$ becomes quadratic in $\vbf_k^{(m)}$ and thus has a
unique solution; the optimal  $\vbf_k^{(m)}$ with maximum SINR  at
low SNR is simply the eigenvector of $\Hbf_{kk}^H\Hbf_{kk}$
associated with the largest eigenvalue. This low SNR behavior is
seen at 0 dB SNR. Fig. \ref{fig:single_channel_N4M4d2} shows the
sum rate performance averaged over all random initializations in
this case. The average behavior is almost the same as that of each
mode; the two-layer optimal beamforming under interference
alignment matches the max-SINR algorithm at high SNR. One
noticeable thing is that the original max-SINR algorithm in
\cite{Gomadam&Jafar:08ArxiV} can yield linearly dependent beam
vectors and does not guarantee the DoF, as shown in the black
curve in Fig. \ref{fig:single_channel_N4M4d2}. (It is easy to show
that linearly dependent beams can be a fixed point of the original
max-SINR algorithm at high SNR by setting
$\vbf_k^{(1)}[n]=e^{j\theta}\vbf_k^{(2)}[n]$ and deriving
$\vbf_k^{(1)}[n+1]$ and $\vbf_k^{(2)}[n+1]$ as in Theorem 2.)
 However, this problem
was fixed easily by inserting orthogonalization after each step.
This insertion does not cause any loss  at high SNR since
outer-coder optimal fixed points with DoF guarantee themselves
have orthonormal columns. Also,  the simulation shows that at low
and intermediate SNR (0 to 20 dB SNR) the algorithm yields the
same performance with or without orthogonalization.

\begin{figure}[htbp]
\centerline{ \SetLabels
\L(0.25*-0.1) (a) \\
\L(0.76*-0.1) (b) \\
\endSetLabels
\leavevmode
\strut\AffixLabels{
\scalefig{0.53}\epsfbox{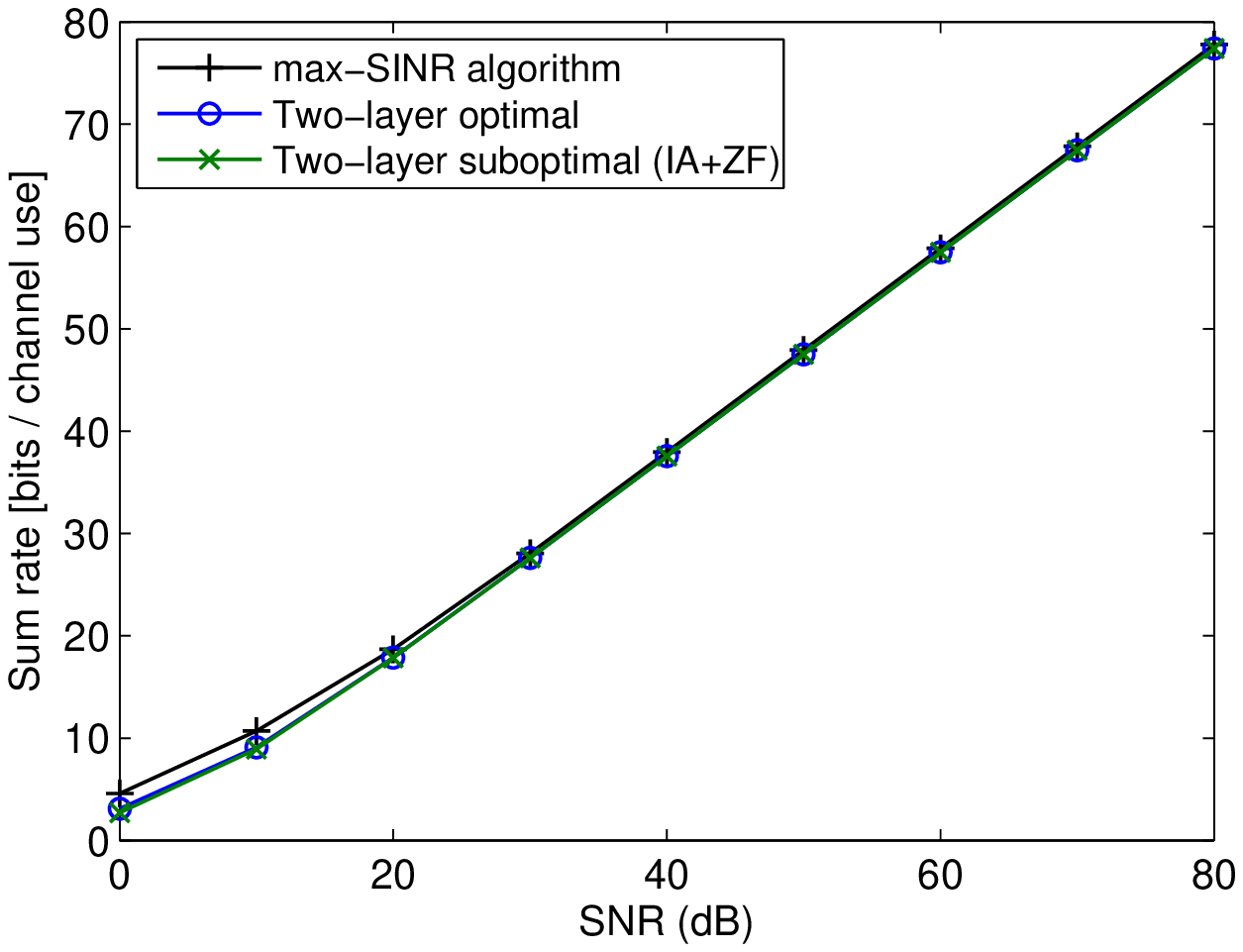}
\scalefig{0.53}\epsfbox{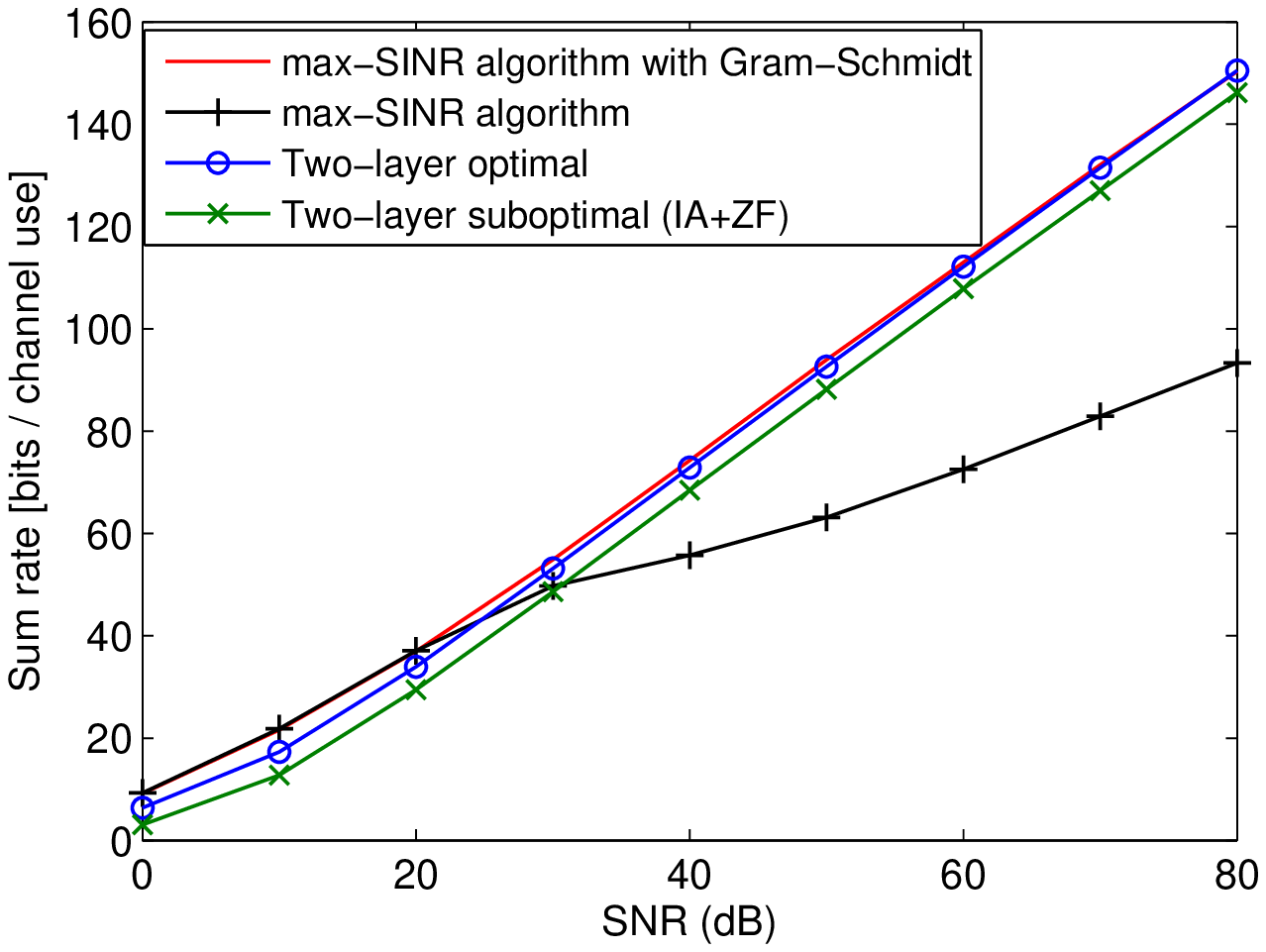} } }
\vspace{1cm} \caption{Sum rate performance averaged over 100
random channel realizations (a) $M=2d=2$ and (b) $M=2d=4$}
\label{fig:multi_channel}
\end{figure}

Finally, Fig. \ref{fig:multi_channel} show the sum rate
performance of the algorithms with one initialization averaged
over 100 different random channel realizations from a Gaussian
distribution. Similar behavior is seen among the three algorithms.
The performance gap of the zero-forcing outer filter and optimal
outer filter under interference alignment can be obtained
similarly to the classical point-to-point MIMO case
\cite{Tse:book} as
\begin{equation}
\Delta C =  K\left(\sum_{i=2}^{d} {\mathbb{E}} \{\log\chi_{2i}^2\}
         -(d-1){\mathbb{E}}\{\log\chi_2^2\} \right),
\end{equation}
for the Gaussian channel, where $\chi_{2i}^2$ denotes the
chi-squared distribution with $2i$ degrees of freedom. For the
example of $d=2$, $\Delta C \approx 4.328$ bits, which matches the
simulation well.

\section{Conclusion}
\label{sec:conclusion}

In this paper we have considered several beamformer design
algorithms for time-invariant MIMO interference channels including
 interference alignment and sum-rate based algorithms such as the
max-SINR and sum-rate gradient algorithms. We have established the
relationship between the sum-rate based algorithms and optimal
linear beamforming under interference alignment given by the
two-layer beamforming structure consisting of inter-user
interference-aligning inner filters and single-user optimal outer
filters. We have shown the optimality of the max-SINR algorithm at
high SNR and the algorithmic advantage of the stream-by-stream
approach, and have also established the structure of the fixed
point set and local convergence of the max-SINR algorithm at high
SNR and the single mode behavior of the algorithm at low SNR. The
optimality here is in the sense that the algorithm contains the
globally optimal beamformer in the fixed point set, and thus
driving the algorithm to the globally optimal fixed point with
arbitrary initialization is still an open issue.


\begin{thebibliography}{}

\bibitem{Cadambe&Jafar:08IT} V. R.~Cadambe and S. A. Jafar,
``Interference alignment and degrees of freedom of the $K$-user interference channel,"
{\it IEEE Transactions on Information Theory}, vol. 54, no. 8, pp. 3425 -- 3441, Aug. 2008.


\bibitem{Bresler&Parekh&Tse:10IT} G. Bresler, A. Parekh, and D. Tse,
``The approximate capacity of the many-to-one and one-to-many Gaussian interference channels,"
{\it IEEE Transactions on Information Theory}, vol. 56, no. 9, pp. 4566 -- 4592, Sep. 2010.


\bibitem{Cadambe&Jafar&Shamai:09IT} V. R.~Cadambe, S. A. Jafar, and S. Shamai,
``Interference alignment on the deterministic channel and application to fully connected Gaussian interference networks," {\it IEEE Transactions on Information Theory}, vol. 55, no. 1, pp. 269 -- 274, Jan. 2009.


\bibitem{Sridharan&Jafarian&Viswanath&Jafar:08Globecom} S. Sridharan, A. Jafarian, S. Vishwanath, and S. A. Jafar, ``Capacity of symmetric $K$-user Gaussian very strong interference channels," in {\it Proceedings of the 2008 IEEE Global Communications Conference}, New Orleans, LA, Dec. 2008.


\bibitem{Sridharan&Jafarian&Vishwanath&Jafar:08Arxiv} S. Sridharan, A. Jafarian, S. Vishwanath, S. A. Jafar, and S. Shamai, ``A layered lattice coding scheme for a class of three user Gaussian interference channels," {\it ArXiv pre-print cs.IT/0809.4316}, Sep. 2008.


\bibitem{He&Yener:09ITW} X. He and A. Yener, ``$K$-user interference channels: Achievable secrecy rate and degrees of freedom," in {\it Proceedings of the IEEE Information Theory Workshop on Networking and Information Theory}, pp. 336 -- 340, Volos, Greece, Jun. 2009.


\bibitem{Etkin&Ordentlich:09Arxiv} R. Erkin and E. Ordentlich, ``On the degrees-of-freedom for the $K$-user Gaussian interference channel," {\it ArXiv pre-print cs.IT/0901.1695}, Jan. 2009.


\bibitem{Motahari&Gharan&MadddahAli&Khandani:09Arxiv2} A. S. Motahari, S. O. Gharan, M. A. Maddah-Ali, and A. K. Khandani, ``Real interference alignment: Exploiting the potential of signal antenna systems," {\it ArXiv pre-print cs.IT/0908.2282v2}, Nov. 2009.


\bibitem{Gou&Jafar:08Arxiv} T. Gou and S. A. Jafar, ``Degrees of freedom of the $K$ user $M \times N$ MIMO interference channel," {\it ArXiv pre-print cs.IT/0809.0099}, Aug. 2008.


\bibitem{Jafar&Shamai:08IT} S. A. Jafar and S. Shamai, ``Degrees of freedom region for the MIMO $X$ channel," {\it IEEE Transactions on Information Theory}, vol. 54, no. 1, pp. 151 -- 170,~ Jan. 2008.


\bibitem{Maddah-Ali&Motahari&Khandani:08IT} M. A. Maddah-Ali, A. S. Motahari, and A. K. Khandani, ``Communication over MIMO $X$ channels: Interference alignment, decomposition, and performance analysis," {\it IEEE Transactions on Information Theory}, vol. 54, no. 8, pp. 3457 -- 3470, Aug. 2008.


\bibitem{Weingarten&Shamai&Kramer:07ITA} H. Weingarten, S. Shamai, and G. Kramer, ``On the compound MIMO broadcast channel," in {\it Proceedings of the Annual Information Theory and Applications Workshop}, La Jolla, CA, Feb. 2007.


\bibitem{Yetis&Gou&Jafar&Kayran:10SP} C. M. Yetis, T. Gou, S. A. Jafar, and A. H. Kayran, ``On feasibility of interference alignment in MIMO interference networks," {\it IEEE Transactions on Signal Processing}, vol. 58, no. 9, pp. 4771 -- 4782, Sep. 2010.


\bibitem{Gomadam&Jafar:08ArxiV} K. Gomadam, V. R.~Cadambe, and S. A. Jafar, ``Approaching the capacity of wireless networks through distributed interference alignment," {\it ArXiv pre-print cs.IT/0803.3816},  Mar. 2008.


\bibitem{Peters&Heath:09ICASSP} S. W. Peters and R. W. Heath, ``Interference alignment via alternating minimization," in {\it Proceedings of the 2009 IEEE International Conference on Acoustics, Speech and Signal Processing},  Taipei, Taiwan, Apr. 2009.


\bibitem{Yu&Sung:10SP} H. Yu and Y. Sung, ``Least squares approach to joint beam design for interference alignment in multiuser multi-input multi-output interference channels," {\it IEEE Transactions on Signal Processing}, vol. 58, no. 9, pp. 4960 -- 4966, Sep. 2010.


\bibitem{Kumar&Xue:10ISIT} K. R. Kumar and F. Xu, ``An iterative algorithm for joint signal and interference alignment," in {\it Proceedings of the 2010 IEEE International Symposium on Information Theory}, Austin, TX, Jun. 2010.


\bibitem{Sung&Park&Lee&Lee:10WCOM} H. Sung, S. Park, K. Lee, I. Lee, ``Linear precoder designs for $K$-user interference channels," {\it IEEE Transactions on Wireless Communications}, vol. 9, no. 1, pp. 291 -- 301, Jan. 2010.


\bibitem{Censor&Zenios:book} Y. Censor and S. A. Zenios, {\it Parallel Optimization: Theory, Algorithms and Applications}. New York:Oxford University Press, 1997.


\bibitem{Poor:book} H.~V.~Poor, {\it An Introduction to Signal Detection and
Estimation, 2nd ed.}, New York: Springer, 1994.


\bibitem{Vishwanath&Jindal&Goldsmith:02ICC} S. Vishwanath, N. Jindal, and A. Goldsmith,  ``On the capacity of multiple input multiple output broadcast channels," in {\it Proceedings of the 2002 IEEE International Conference on Communications}, vol. 3, pp. 1444 -- 1450, New York, Aug, 2002.


\bibitem{Jindal&Vishwanath&Goldsmith:04IT} N. Jindal, S. Vishwanath, and A. Goldsmith,  ``On the duality of Gaussian multiple-access and broadcast channels," {\it IEEE Transactions on Information Theory}, vol. 50, no. 5, pp. 768 -- 783, May, 2004.


\bibitem{Viswanath&Tse:03IT} P. Viswanath and D. Tse,  ``Sum capacity of the vector Gaussian broadcast channel and uplink-downlink duality," {\it IEEE Transactions on Information Theory}, vol. 49, no. 8, pp. 1912 -- 1921, Aug, 2003.


\bibitem{Magnus&Neudecker:book} J. R. Magnus and H. Neudecker, {\it Matrix Differential Calculus with Applications in Statistics and Economics, revised ed.}, New York: John Wiley \& Sons, 1999.


\bibitem{Brandwood:83IEEproc} D. H. Brandwood, ``A complex gradient operator and its application in adaptive array theory," {\it IEE Proceedings H Microwaves, Optics and Antennas}, vol. 130, no. 1, pp. 11 -- 16, Feb. 1983.


\bibitem{Telatar:99} \.{I}. Telatar, ``Capacity of multi-antenna Gaussian channels," {\it European Transactions on Telecommunications}, vol. 10, no. 6, pp. 585 -- 596, Nov.-Dec. 1999.


\bibitem{Stark&Yang:book} H. Stark and Y. Yang, {\it Vector Space Projections}, New York: Wiley, 1998.


\bibitem{Tse:book} D. Tse and P. Viswanath, {\it Fundamentals of Wireless Communication}, Cambridge, UK: Cambridge University Press, 2005.


\end{thebibliography}

\end{document}
